\documentclass[format=acmsmall, review=false]{acmart}

\usepackage{booktabs}

\let\chapter\section
\usepackage[ruled,vlined]{algorithm2e}
\usepackage{mathrsfs}
\usepackage{graphicx}
\usepackage{amsmath}
\usepackage{amssymb}
\usepackage{multirow} 
\usepackage{wrapfig}

\usepackage[position=top]{subfig}
\usepackage{tikz}
\usetikzlibrary{arrows,decorations,decorations.shapes,decorations.markings,decorations.pathreplacing,backgrounds,shapes,petri,topaths}
\usepackage{tkz-berge}
\usepackage{pgfplots}
\usepackage{multirow}
\usepackage{enumitem}
\newtheorem{observation}[theorem]{Observation}
\DeclareMathOperator{\argmin}{argmin}

\newenvironment{rtheorem}[1]{\medskip\textsc{Theorem~\ref{#1}.}\begin{itshape}}{\end{itshape}}

\newenvironment{rproposition}[1]{\medskip\textsc{Proposition~\ref{#1}.}\begin{itshape}}{\end{itshape}}
\newenvironment{rcorollary}[1]{\medskip\textsc{Corollary~\ref{#1}.}\begin{itshape}}{\end{itshape}}

\newcommand{\hatv}[1]{\hat{\vec #1}}
\def\shortcite{\cite}
\def\mock{}

\newcommand{\eps}{\epsilon}
\renewcommand{\vec}[1]{\mathbf{#1}}
\def\GBC{\mathcal{G}}

\def\cal{\mathcal}
\newcommand{\coursename}{(67686) Mathematical Foundations of AI}

\newcommand{\handout}[5]{
   \renewcommand{\thepage}{#1-\arabic{page}}
   \noindent
   \begin{center}
   \framebox{
      \vbox{
    \hbox to 5.78in { {\bf \coursename}
         \hfill #2 }
       \vspace{4mm}
       \hbox to 5.78in { {\Large \hfill #5  \hfill} }
       \vspace{2mm}
       \hbox to 5.78in { {\it #3 \hfill #4} }
      }
   }
   \end{center}
   \vspace*{4mm}
}

\def\mock{\!\!\!\!\phantom{\frac{1}{1}}}

\def\({\left(}
\def\){\right)}

\newcommand{\labeq}[2]{
\begin{equation}
\label{eq:#1} #2
\end{equation}}
\def\ol{\overline}
\def\ul{\underline}

\def\PoA{\mbox{PoA}}
\def\BPoA{\mbox{BPoA}}

\newenvironment{proof-sketch}{\noindent{\bf Sketch of Proof}\hspace*{1em}}{\qed\bigskip}
\newenvironment{proof-idea}{\noindent{\bf Proof Idea}\hspace*{1em}}{\qed\bigskip}
\newenvironment{proof-of-lemma}[1]{\noindent{\bf Proof of Lemma #1}\hspace*{1em}}{\qed\bigskip}
\newenvironment{proof-attempt}{\noindent{\bf Proof Attempt}\hspace*{1em}}{\qed\bigskip}

\makeatletter
\def\fnum@figure{{\bf Figure \thefigure}}
\def\fnum@table{{\bf Table \thetable}}
\long\def\@mycaption#1[#2]#3{\addcontentsline{\csname
  ext@#1\endcsname}{#1}{\protect\numberline{\csname
  the#1\endcsname}{\ignorespaces #2}}\par
  \begingroup
    \@parboxrestore
    \small
    \@makecaption{\csname fnum@#1\endcsname}{\ignorespaces #3}\par
  \endgroup}
\def\mycaption{\refstepcounter\@captype \@dblarg{\@mycaption\@captype}}
\makeatother

\newcommand{\mathify}[1]{\ifmmode{#1}\else\mbox{$#1$}\fi}
\newcommand{\bigO}O

\newcommand\stup[1]{\langle #1 \rangle}
\newcommand\tup[1]{\left\langle #1 \right\rangle}

\def\PW{\mbox{SPW}}

\renewcommand{\vec}[1]{\boldsymbol{#1}}

\newcommand{\calC}{{\cal C}}

\newcommand{\remove}[1]{{}}

\def\GG{\mathbf{G}}
\newcommand{\xqed}{\mbox{\raggedright $\Diamond$}}

\newcommand{\newpar}[1]{
\vspace{-0mm}
\paragraph{#1}}

\newcommand{\omittext}[1]{}

\newcount\Comments  
\definecolor{darkgreen}{rgb}{0,0.6,0}
\newcommand{\kibitz}[2]{\ifnum\Comments=1{\color{#1}{#2}}\fi}
\newcommand{\rmr}[1]{\kibitz{blue}{[RESHEF:#1]}}

\def\ol{\overline}

\begin{document}

\title{Playing the Wrong Game: Bounding  Externalities in Diverse Populations of Agents}
\author{Reshef Meir}
\affiliation{%
  \institution{Technion---Israel Institute of Technology}
  \city{Haifa}
  \postcode{32000}
  \country{Israel}}

\author{David Parkes}
\affiliation{
\institution{Harvard University}
\city{Cambridge}
\postcode{02138}
\state{MA}
\country{USA}
}


\begin{abstract}
	The robustness of multiagent systems can be affected by mistakes or behavioral biases (e.g., risk-aversion, altruism, toll-sensitivity), with some agents playing the ``wrong game.'' This can change the set of equilibria, and may in turn harm or improve the social welfare of agents in the system. We are interested in bounding what we call the {\em biased price of anarchy} (BPoA) in populations with diverse agent behaviors, which is the ratio between welfare in the ``wrong'' equilibrium and optimal welfare.  
	We study nonatomic routing games, and derive an externality bound that depends on a key topological parameter of the underlying network. 
	We then prove two general BPoA bounds for games with diverse populations: one that relies on the network structure and the \emph{average bias} of all agents in the population, and one that is independent of the structure but depends on the \emph{maximal bias}. Both types of bounds can be combined with known results to derive concrete BPoA bounds for a variety of specific behaviors (e.g., varied levels of risk-aversion). 
	\end{abstract}
%
%
%
%
%

\renewcommand{\labelitemi}{$\bullet$}
\renewcommand{\labelitemii}{$\circ$}

\maketitle

\section{Introduction}
Game theory is founded on the assumption that agents are rational
decision makers, i.e. maximizing their utility, and that groups of
agents reach an equilibrium outcome. 
In many games there is some objective measure of welfare that can be accurately measured in terms of money, time, and so on. Utility on the other hand, is subjective. It is not always easy to identify an explicit utility function that an agent is trying to maximize, if such exists. Even when there are objective factors that affect agents' utility, such as profit, effort, uncertainty, and temporal delays, various agents may weigh these factors differently or have subjective constraints and biases. Thus different agents may demonstrate different behavior even in the same situation. 
%

As a concrete example, commuters may have some information on the expected congestion at each route via traffic reports or a cellphone app. However they also know that this information is inaccurate, and a risk-averse driver might take into account not just the expected congestion, but also the likelihood of an unexpected delay, a heuristic safety margin and so on. Moreover, different commuters may have different levels of risk-aversion,  or act upon different heuristics. 

The implications of these subjective differences and biases
on a multiagent system are two-fold.
First, from the perspective of an outside observer who cares about a particular objective (say, total latency), the agents are
playing the ``wrong game.'' This is true whether they are failing to optimize their true cost due to cognitive limitations and biases~\cite{Devetag08}, or optimizing their true subjective costs that differ from those of the ``objective'' observer.

A second related issue, is that  the behavior of one agent may exert significant negative externality on the utility of another. The extreme case is when some agents are adversarial, and act in a way that tries to \emph{minimize} the utility of some other agents.

It is well known that even in games without biases or subjective utilities, there may be negative externalities, and that equilibrium outcomes may be suboptimal in terms of the utilitarian social welfare. This inefficiency is often quantified as the \emph{Price of Anarchy} (PoA), which is the ratio between the  social welfare at the worst equilibrium and the optimal social welfare. Biases may change the equilibria of the game, and thus have a large effect on the PoA.  
Externalities are widely studied in multiagent systems, but usually in the context of well-defined behavior such as maximizing utility in a game~\cite{krysta2010combinatorial,oh2008few} or finding an optimal or stable solution~\cite{branzei2013matchings,caire2011conviviality,banerjee2010coalition}. How can we hope to bound externalities in a diverse society of agents with assorted biases and behaviors?   The answer will lie in a proper abstraction of such behaviors, but before that we will present our questions more formally.

 \newpar{Research goals} Suppose
that in game $\GG$, each agent $i$ has some true cost 
function $c^i$ (in this paper we consider negative utilities, i.e. costs). Now, each agent $i$ sees her cost as some other
function $\hat c^i$, and thus we are interested in the equilibria of
the {\em biased game} $\hat \GG$ comprised of modified utilities or costs
$\{\hat{c}^i\}_{i=1,2,\dots}$. We ask the following: 
\begin{itemize}[topsep=0.5ex,itemsep=0.1ex,leftmargin=0.3cm] 
	\item Is there a bound to the negative externality that type~$j$ exerts on type~$i$, that applies \emph{regardless} of the type $\hat c^j$? (Example: can we bound the equilibrium latency of type $i$ commuters without any assumption on the risk-aversion level of others?)
\item Is there a bound to the social cost (in the ``objective game'' $\GG$) of $\hat \GG$'s worst equilibrium? (Example: can we bound the total latency in equilibrium, given all commuters' subjective cost functions?)
\end{itemize}
We refer to the ratio between the latter measure and the optimal social cost in $\GG$ as the \emph{Biased Price of Anarchy} (BPoA), and note that it coincides with the PoA when $\hat \GG=\GG$.  

While the first question is straight-forward, the second question may raise some conceptual debate. One  may argue that since agents are acting so as to minimize their biased cost $\hat c^i$ rather than $c^i$, this is the cost we should take into account when computing the social cost or welfare. We justify using $\GG$ as the baseline for welfare
as follows.
First, the social cost may be the objective used by the system designer or analyst, as in our example above. A different analyst might care about a different goal, resulting in a different objective game $\GG$.
Second, the social cost may be the sum of the agents' true utilities, 
while the agents are bounded-rational. For example they may be unaware of some roads~\cite{acemoglu2016informational} or  the exact latency functions~\cite{tennenholtz2009learning}; they may have uncertainty regarding actual congestion~\cite{vasserman2015implementing} or the overall amount of other agents~\cite{ashlagi2009two};  and may assign wrong probabilities to rare outcomes~\cite{hertwig2004decisions}.

\newpar{Smoothness in routing games}
Nonatomic routing games in the  Wardrop
model~\cite{wardrop1952road} are a good testbed for the questions above: they have
very convenient theoretical properties, such as the uniqueness of equilibrium (up to identical
utilities); equilibrium inefficiency and in particular the PoA is very well understood; and several
 biases have already been suggested and
studied in the context of routing games (see Related Work).
The \emph{smoothness method} allows to leverage a property of the edge cost functions to obtain a
tight upper bound on the Price of Anarchy that is independent of the network
topology: if all cost functions are $(\lambda,\mu)$-smooth for some parameters $\lambda,\mu$, then the PoA is bounded by $\frac{\lambda}{1-\mu}$~\cite{roughgarden2003price,correa2008geometric,R09}.
Thus the smoothness of cost functions lets us \emph{abstract away} the details of the game and prove PoA bounds on large classes of games.

\rmr{
In this paper,  we focus on nonatomic routing games and study the
important \emph{heterogeneous} setting where participants can each
have different, arbitrary biases.
Our goal is to leverage 
general properties such as known PoA bounds and smoothness bounds, in order to derive BPoA bounds for heterogeneous populations---much like standard smoothness is leveraged to derive PoA bounds for broad classes of games. This would enable us to take results (both known and new) from simpler games, and combine them  to bound the objective total latency in games with arbitrary heterogeneous biases.
%
%
This approach contrasts with the few 
known results involving heterogeneous populations (see Related Work),
which are highly specific both in the focus on a certain bias, and in that they add restrictions on the network topology (parallel or series-parallel).
}

\subsection{Paper structure and contribution}

After a short overview of nonatomic routing games, price of anarchy, and a measure of graph complexity called \emph{serial-parallel width} (Section~\ref{sec:prelim}),  we  prove a tight bound on the negative externality in any routing game without any assumption on agents' behavior (Section~\ref{sec:ex}). Our bound generalizes previous results from specific behaviors on series-parallel networks~\cite{NS15,acemoglu2016informational} to arbitrary behaviors and networks (parametrized by their serial-parallel width).

Next, we adopt \emph{smoothness} as  an abstraction for general biases and behaviors. 
In
Section~\ref{sec:bias} we extend the definition of a smooth cost
function to $(\hat \lambda,\hat \mu)$-biased-smoothness that takes into account  both the ``true'' and the ``biased'' cost functions.  This approach follows similar extensions for specific, modified
costs~\cite{bonifaci2011efficiency,chen2011robust}, and we review recent smoothness bounds for such specific settings.  
%

In Section~\ref{sec:diverse}, we consider games where agents have diverse biases, and use our results from Section~\ref{sec:ex} along with biased-smoothness to derive several bounds on the Biased Price of Anarchy. 
For example, for symmetric games over series-parallel networks (which have parallel-width of $1$), we derive a BPoA bound  in terms of the ``average'' biased-smoothness: $\BPoA \leq O(1)\sum_i \frac{r_i}{r}\frac{\hat \lambda^i}{1-\hat \mu^i}$, where $\frac{r_i}{r}$ is the fraction of type~$i$ agents in the population.  For arbitrary networks, we get a bi-criteria result that depends both on the average bias as above, and on the serial-parallel width of the underlying network.


For the special case of polynomial cost functions, we leverage (Section~\ref{sec:Psi}) known results on the PoA in heterogeneous unbiased games to derive a structure-independent BPoA bound. In contrast to our main results, this bound depends on the \emph{worst} bias rather than on the average bias. 
%

	Appendices \ref{sec:structure} and \ref{sec:no_structure} contain proofs for Sections \ref{sec:diverse} and \ref{sec:Psi}, respectively. Appendix~\ref{sec:biases} contains a detailed smoothness analysis of various cost functions. Appendix~\ref{sec:PoA} presents some bounds on the price of anarchy in (unbiased) player-specific games that can be derived from our results. 

\subsection{Related Work}
\label{sec:related}


The most-well studied, modified cost function
comes about as a result of
tolls, where the travel time plus the imposed toll can
be thought of as a modified cost
function. 
In this context, most papers focus on the objective of minimizing total latency~\cite{cole2003pricing,fleischer2004tolls,kleer2016impact}, and on the design of optimal or practical toll schemes~\cite{bonifaci2011efficiency,jelinek2014computing,fotakis2015improving,sharon2016delta}. Heterogeneous biases arise when different agents have different sensitivity to imposed tolls.
 We explain in the relevant sections how these papers technically relate to our work.

%
%
%
%
%
Chen et al.~\shortcite{chen2011robust,chen2014altruism} apply smoothness analysis to
provide BPoA bounds for various games (including atomic congestion
games) where agents are altruistic, i.e., part of their utility is
derived from the social welfare. In the context of nonatomic routing games, their model is formally equivalent to toll-sensitivity (see Section~\ref{sec:h_bounds}).


Acemoglu et al.~\shortcite{acemoglu2016informational} study nonatomic congestion games where some agents are unaware of the existence of certain edges, which is equivalent to having a wrong cost function that assigns infinite costs to some edges. They prove that on directed series-parallel networks, such ignorance can only lead to a worse equilibrium than under true information, yet the \emph{worst-case} PoA remains the same.


Another behavioral bias that has been  studied in congestion
games is risk
aversion~\cite{ashlagi2009two,angelidakis2013stochastic,piliouras2016risk,NS15,MP15}, which can often be written as biased cost functions
(see Section~\ref{sec:h_bounds} for details). 


Finally, Babaioff et al.~\shortcite{babaioff2007congestion} consider
congestion games where some of the agents are malicious. Babaioff et al. focused on the effect of a negligible amount of malicious agents, and showed examples where it can be either detrimental or (surprisingly) beneficial to the other agents, but without any upper bounds.  
Indeed, malicious behavior can be considered as another form of bias. 


 %

        \section{Preliminaries}

        
\label{sec:prelim}
A \emph{network} is a 2-terminal directed multigraph $G=\tup{V,E,s,t}$, where $s,t\in V$ are special vertices (\emph{source} and \emph{target}), and every edge $e\in E$ belongs to some simple $s-t$ path. 

\begin{wrapfigure}[7]{r}[-0.4\width]{0.08\textwidth}
\vspace{-4mm}
\begin{tikzpicture}[scale=0.55,transform shape]

  \Vertex[L=$a$,x=2.3,y=1.2]{a}
  \Vertex[L=$s_1$,x=2,y=2.4]{s1}
  \Vertex[L=$t_1$,x=2,y=0]{t1}
	\Vertex[x=1,y=1.8]{b}
	\Vertex[x=1,y=0.6]{c}
	
	\Vertex[L=$s_2$,x=4,y=2.4]{s2}
	\Vertex[L=$t_2$,x=4,y=0]{t2}
	\Vertex[x=4,y=1.2]{d}
  \tikzstyle{VertexStyle}=[fill=black!20!white]
  
  \tikzstyle{LabelStyle}=[fill=white,sloped]
  \tikzstyle{EdgeStyle}=[->]
  \Edge(a)(t1)
  \Edge(s1)(b)
  \Edge(b)(c)
  \Edge(c)(t1)
  
	\tikzstyle{EdgeStyle}=[->,bend right]
	\Edge(s1)(a)
	\Edge(s2)(d)
	\Edge(d)(t2)

	\tikzstyle{EdgeStyle}=[->,bend left]
	\Edge(s1)(a)
	\Edge(s2)(d)
	\Edge(d)(t2)

	\tikzstyle{EdgeStyle}=[-,dotted]
	\Edge(s1)(s2)
	\Edge(t1)(t2)
\end{tikzpicture}

\vspace{-6mm}
\end{wrapfigure}

A network is \emph{series-parallel}~\cite{eppstein1992parallel,jakoby2006space} if it is either a single edge, or composed recursively by joining two series-parallel networks in series  or in parallel. E.g., merging $\{s_1,s_2\}$ and $\{t_1,t_2\}$ in the following networks also results in a series-parallel network:

\subsection{Nonatomic routing games}\label{sec:NRG}

Following the definitions of Roughgarden~\shortcite{roughgarden2003price} and
Roughgarden and Tardos~\shortcite{roughgarden2004bounding}, a {\em nonatomic routing game} (NRG)
 is a tuple $\GG = \tup{G,m,\vec c,\vec s,\vec t,\vec r}$,  where 
\begin{itemize}
	\item $G=\tup{V,E}$ is a directed multigraph;
	\item $m\in \mathbb N$ is the number of agent types;
	\item $\vec c = (c_e)_{e\in E}$, where $c_e(x)\geq 0$ is the cost incurred when $x$ agents use edge $e$;\footnote{We use the term cost rather than latency, to reflect that agents may care about other factors.}
	\item $\vec s,\vec t\in V^m$, where $(s_i,t_i)$ are the source and target nodes of type $i$ agents;
	\item $\vec r=(r_i)_{i\leq m}$, where $r_i > 0$ is the total mass of type~$i$ agents. 
 The total mass of agents of all types is $\sum_{i\leq m} r_i=r$, and we assume unless specified otherwise that $r=1$.
\end{itemize}

We denote by  $A_i \subseteq 2^E$ the set of all directed simple paths between the pair of nodes $(s_i,t_i)$ in the graph. Thus $A_i$ is the set of {\em actions} available to agents of type $i$. We denote by $A = \cup_i A_i$ the set of all directed source-target simple paths. We assume that all cost functions mentioned in the paper  (including biased costs mentioned later on)  are non-decreasing, continuous, differentiable and semi-convex (i.e., $xc_e(x)$ is convex). Such cost functions are called \emph{standard}~\cite{roughgarden2003price}.

\paragraph{Player-specific costs} 
A \emph{nonatomic routing game with player-specific costs} (PNRG) is a tuple $\GG = \tup{G,m,(\vec c^i)_{i\leq m},\vec s,\vec t,\vec r}$. The difference
from a NRG is that agents of each type $i$ experience a cost of $c^i_e(x)$ when $x$ agents use edge $e$. We can have multiple types with the same source and target nodes to allow diversity of behavior. To avoid confusion, we refer to  $(s_i,t_i)$ (or $A_i$) as the \emph{demand type} and to $\vec c^i$ as the \emph{cost type}.
Thus the type $i$ specifies both the demand type and the cost type.

A PNRG is \emph{symmetric} if all agents have the same demand type, i.e., $A_i=A$ for all $i$.
A PNRG is a \emph{resource selection game} (RSG) if $G$ is a network of parallel links. That is, if the action of every agent is to select a single $(s,t)$ edge.

\newpar{Flows} A \emph{flow} (or action profile) of a PNRG is a vector $\vec f \in \mathbb R_+^{|A| \times m}$,
where $f_{p,i}$ is the amount of agents of type $i$ that use path
$p\in A_i$. In a valid flow, $\sum_{p\in A_i}f_{p,i} = r_i$ for all $i$. The
total traffic on path $p\in A$ is denoted by $f_p = \sum_{i=1}^m
f_{p,i}$. Similarly, the total traffic on edge $e\in E$ is denoted by  $f_e = \sum_{p:e\in p}f_p$.
 Denote the {\em support of type~$i$ strategy} in flow $\vec f$ by $P_i(\vec f) = \{p\in
A : f_{i,p} > 0\}$. That is, all paths used by type~$i$ agents in flow $\vec f$. 

The cost for an agent of type $i$ in game $\GG$, selecting a path $p\in A_i$ in flow $\vec f$, is $c^i(p,\vec f) =  \sum_{e\in p}c^i_e(f_e)$.
\newpar{Social cost}For an NRG (not player-specific), 
we denote by $\mathit{SC}_i(\GG,\vec f)=\sum_{p\in A_i} f_{i,p} c(p,\vec f)$ the cost experienced by type~$i$ agents in flow $\vec f$. 
By summing over all types, we get the social cost:

\begin{align*}
 \mathit{SC}(\GG,\vec f)=&\sum_{i\leq m} \mathit{SC}_i(\GG,\vec f) =\sum_{i\leq m} \sum_{p\in P_i(\vec f)} f_{i,p} c(p,\vec f) 
= \sum_{e\in E}  c_e(f_e) f_e \label{eq:SC_total}
\end{align*}

%

Thus the  social cost only depends on the total traffic per edge. 
%
  We denote by $\vec f^o(\GG)\in \argmin_{\vec f} \mathit{SC}(\GG,\vec f)$ some profile with minimal total cost, and $\mathit{OPT}(\GG)=SC(\GG,\vec f^o(\GG))=\min_{\vec f} \mathit{SC}(\GG,\vec f)$.

\newpar{Equilibrium}
A flow $\vec f$ for an PNRG is an
\emph{equilibrium} in game $\GG$ if for every agent type $i$, any used path
$p \in P_i(\vec f)$ and any $p' \in A_i$, we have  $c^i(p,\vec f) \leq
c^i(p',\vec f)$. That is, if no agent can switch to a path with a lower
cost.  This provides the analogy of a Nash equilibrium for
nonatomic games. 

It is known that in any NRG there is at least one equilibrium, and that this can be reached by a simple best-response dynamic. Further, all equilibria have the same social cost and in every equilibrium all agents of type~$i$ experience the same cost~\cite{
aashtiani1981equilibria,milchtaich2000generic,roughgarden2004bounding,blum2006routing}. 
Player-specific NRGs are also guaranteed to have at least one equilibrium~\cite{schmeidler1973equilibrium}, however, equilibrium costs may not be unique, and best-response dynamics may not converge, except in special cases~\cite{gairing2011routing}.

%
%
%
%
\newpar{Affine routing games}
In an {\em affine} NRG, all cost functions take the form of a linear function. That is, $c_e(x) = a_e x + b_e$ for some constants $a_e\geq 0,b_e\geq 0$.  The social cost can be written as $\mathit{SC}(\GG,\vec f)= \sum_{e\in E} a_e (f_e)^2 + b_e f_e$.
%
{\em Pigou's example} is the special case of an affine RSG with two
resources, where $c_1(x)=1$ and $c_2(x)=ax$ .
We  denote by $\GG_P(a)$ the
instance where $c_2(x) = a x$ (see Fig.~\ref{sfig:Pigou}). 

%
%
%
%

\paragraph{The price of anarchy} 

Let $EQ(\GG)$ be the set of equilibria in game $\GG$.  The \emph{price
  of anarchy} (PoA) of a game is the ratio between the social cost in
the worst equilibrium in $EQ(\GG)$ and the optimal social cost~\cite{koutsoupias1999worst}. Formally, 
%
$\PoA(\GG) =\sup_{\vec f^*\in EQ(G)}\frac{\mathit{SC}(\GG,\vec f^*)}{\mathit{OPT}(\GG)}$. 
For example in affine NRGs, it is known that  $\PoA(\GG)\leq \frac{4}{3}$, and this bound is attained by the Pigou example of $\GG_P(1)$~\cite{roughgarden2004bounding}. 


\paragraph{Smoothness}
A cost function $c$ is $(\lambda,\mu)$-smooth for $\lambda\geq 0, \mu<1$ if for any $x,x'\geq 0$, it holds that 

\labeq{smooth}{c(x)x'\leq \lambda x'c(x')  + \mu x c(x).}

A NRG $\GG$ is $(\lambda,\mu)$-smooth if all cost functions in $\GG$ are $(\lambda,\mu)$-smooth. For any $(\lambda,\mu)$-smooth NRG, $\PoA(\GG)\leq \frac{\lambda }{1-\mu}$~\cite{correa2008geometric,R09}.\footnote{The mentioned papers only require each $c_e$ to be smooth around the equilibrium flow $x=f^*_e$. Our proofs require the cost functions to be smooth at several different places, and hence the slight difference in the definition.}   Moreover, w.l.o.g. $\lambda=1$ (that is, for any class of cost functions there is an optimal pair $(1,\mu)$ for some $\mu$~\cite{correa2004selfish,roughgarden2004bounding}). 
For example, affine functions are $(1,\frac14)$-smooth, which again entails a PoA bound of $\frac43$.

\subsection{Serial-parallel Width}\label{sec:PW}
\begin{figure}
\centering
\subfloat[$\GG_P(a)$]{ \label{sfig:Pigou}
\begin{tikzpicture}[scale=0.75,transform shape]
  \Vertex[x=1.5,y=5]{s}
  \Vertex[x=1.5,y=1]{t}
 \tikzstyle{EdgeStyle}=[->,bend right]
\Edge[label=$1$](s)(t)
 \tikzstyle{EdgeStyle}=[->,bend left]
\Edge[label=$ax$](s)(t)
\end{tikzpicture}
}
\quad\quad\quad
\subfloat[$\GG$]{ \label{sfig:Braess}
\begin{tikzpicture}[scale=0.75,transform shape]

  \Vertex[x=0.4,y=3]{a}
  \Vertex[x=2.6,y=3]{b}
  \Vertex[x=1.5,y=5]{s}
  \Vertex[x=1.5,y=1]{t}
  \tikzstyle{VertexStyle}=[fill=black!20!white]
  
  \tikzstyle{LabelStyle}=[fill=white,sloped]
  \tikzstyle{EdgeStyle}=[->]
  \Edge[label=$x^2$](s)(a)
  \Edge[label=$1$](s)(b)
  \Edge[label=$1$](a)(t)
  \Edge[label=$x^2$](b)(t)
  \Edge[label=$0$](a)(b)
\end{tikzpicture}
}
\quad
\subfloat[$\hat \GG^\alpha$ for $\alpha=3$]{ \label{sfig:Braess_r}
\begin{tikzpicture}[scale=0.75,transform shape]

  \Vertex[x=0.4,y=3]{a}
  \Vertex[x=2.6,y=3]{b}
  \Vertex[x=1.5,y=5]{s}
  \Vertex[x=1.5,y=1]{t}
  \tikzstyle{VertexStyle}=[fill=black!20!white]
  
  \tikzstyle{LabelStyle}=[fill=white,sloped]
  \tikzstyle{EdgeStyle}=[->]
  \Edge[label=$9x^2$](s)(a)
  \Edge[label=$1$](s)(b)
  \Edge[label=$1$](a)(t)
  \Edge[label=$9x^2$](b)(t)
  \Edge[label=$0$](a)(b)
\end{tikzpicture}
}
%
\vspace{-2mm}
\caption{\label{fig:Braess}\eqref{sfig:Pigou} Pigou's example.  \eqref{sfig:Braess} An objective game $\GG$ (a quadratic variation of the Braess paradox). 
 \eqref{sfig:Braess_r} The game $\hat \GG^{\alpha}$  is the same game played by pessimistic agents with parameter $\alpha=3$.
The biased costs of edges with fixed costs like $sb$ do not change, but the biased cost on the edge $sa$ for example is $\hat c_{sa}^\alpha(x)= c_{sa}(3x)=(3x)^2=9x^2$.\vspace{-1mm}}
\end{figure}
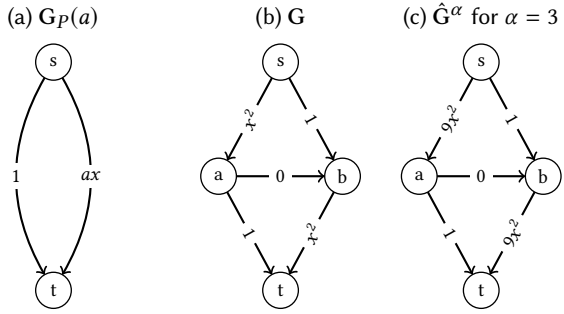

Consider a network $G=\tup{V,E,s,t}$.
\begin{definition}\label{def:PW}
A set of edges $S\subseteq E$  is \emph{parallel} if there is some $S' \subseteq E$ s.t. $S\subseteq S'$, and $S'$ is a minimal cut between $s$ and $t$ in the network $G$.    
	\end{definition}
\begin{definition}\label{def:crossed_edges}
 A set of edges $S\subseteq E$ is \emph{serial} if there is a simple directed $s-t$ path $p$ containing $S$. 
\end{definition}
\begin{definition}[Serial-parallel Width]
The \emph{serial-parallel width} of a network, $\PW(G)$, is the size of the largest set $S\subseteq E$ that is both parallel and serial. 
\end{definition}
Intuitively, a serial-parallel width of $k$ means there are at least $k$ non-intersecting source-target paths, and some additional path that edge-intersects all of them.
\begin{example}
Consider the Braess network  in Fig.~\ref{sfig:Braess} (ignore the costs). The minimal $s-t$ cuts  are: $\{sa,sb\}, \{at,bt\}$,  $\{sa,bt\}$ and $\{sb,ab,at\}$. Thus the set $\{sa,bt\}$ is both parallel and serial, which means $\PW(G_B)\geq 2$. The set $\{sa,at\}$ is serial but not parallel; and  $\{sa,sb,ab\}$ is neither. In fact, the \emph{only} parallel set of size greater than $2$ is $\{sb,ab,at\}$, which is not serial, thus $\PW(G_B)<3$. We conclude that the serial-parallel width of the Braess network is $2$.
\end{example}

\begin{definition}\label{def:CP_k}
For any $k\geq 2$, we define the \emph{$k$-serial-parallel graph} $G_{SP(k)}$ as follows. $G=\tup{V,E,s,t}$, where  $V=\{s,t,a_2,\ldots,a_k,b_1,\ldots,b_{k-1}\}$, and $E=\bigcup_{i=2}^{k-1}\{(s,a_i),(a_i,b_i),(b_i,t),(b_i,a_{i+1})\} \cup \{(s,b_1),(a_k,t)\}$. 
\end{definition}

See  Figure~\ref{fig:CPk} for an example. 
%

The $2$-serial-parallel network is the Braess network $G_B$. 
The serial-parallel width of the $k$-serial-parallel network is  exactly $k$, where  the parallel edges are $\{(s,b_1),(a_2,b_2),\ldots,(a_{k-1},b_{k-1}),(a_k,t)\}$.
The network $G_{SP(k)}$ was used in \cite{babaioff2007congestion} to derive examples of games with high Price of Malice, and we will use it later in a similar way. 
In a companion paper, Meir and Deligkas~\shortcite{meir17embedded} proved that the serial-parallel width of acyclic networks can be characterized completely, where $\PW(G)<k$ if and only if $G_{SP(k)}$ (or some small variants of which) is not embedded in $G$. For $k=1$, this entails another characterization.  
%
\begin{proposition}[Meir and Deligkas~\cite{meir17embedded}]
\label{th:PW_SP}
Let $G$ be an acyclic network, then $\PW(G)=1$ if and only if $G$ is series-parallel.
\end{proposition}

\section{Bounding Externalities}\label{sec:ex}
Suppose we are given a game $\GG=\stup{G,m,({\vec c}^j)_{j\leq m},\vec s,\vec t,\vec r}$, 
and agents play some equilibrium $\vec f^*$ of $\GG$. For any single type $i$,  all type~$i$ agents have the same cost in $\vec f^*$. We denote this cost by $C^{i,*}=c^i(p,\vec f^*)$, where $p\in P_i(\vec f^*)$ is an arbitrary path used by a type~$i$ agent. Note that $r_i \cdot C^{i,*}=SC_i(\GG,\vec f^*)$.

Our goal in this section is to bound $C^{i,*}$. A-priori, this may seem difficult, as we do not assume \emph{anything} about the types of the other agents. However, we will show that the negative externality of the other types can be bounded using the structural parameters of the network $G$ (the serial-parallel width), and even adversarial agents cannot be much worse for $i$ than more type~$i$ agents.

Given a game $\GG$ and type $i$, we define a new game $\GG^i$ by setting both the cost type and the demand type of all agents in $\GG$ to $i$. That is, $\GG^i=\stup{G,1,\vec c_i,s_i,t_i, r}$.
We also define a  game $\GG^{(k)}=\stup{G,m,\vec c,\vec s,\vec t,k\cdot r}$, where the demand in $\GG$ is multiplied by $k$. Finally, we set $\GG^{i,(k)}=(\GG^i)^{(k)}$, i.e., $\GG^{i,(k)}$ is a game where there all $k\cdot r$ agents are of type~$i$. Note that $\GG^{i,(1)} = \GG^i$, and denote $C^i=C^{i,1}$. 

 Let $\vec g^*$ be an equilibrium of $\GG^{i,(k)}$. As $\GG^{i,(k)}$ is a symmetric game, all agents have the same cost in $\vec g^*$.  We denote this cost by $C^{i,(k)}=c^i(p,\vec g^*)$, where $p\in \ol P(\vec g^*)$ is an arbitrary used path.

Finally, let $G|_i=\tup{V^i,E^i,s_i,t_i}$ be the network obtained from $G=\tup{V,E}$, by eliminating all edges and vertices that are not in $A_i$.

The following bound is our main result in this section. Prop.~\ref{th:CPk_tight} shows that the bound is tight. 
\begin{theorem}[Externality theorem] \label{th:PW_bound} Let $\GG$ be any PNRG played on a network $G$. If $\PW(G|_i)\leq k$, then $C^{i,*} \leq C^{i,(k)}$.
\end{theorem}
\rmr{can we improve the bound to take into account $1-r_i$ ?  that is, for small $\eps=1-r_i$ we would expect a small influence.} 
Before we prove the theorem, we explain its implications. 
For $k=1$ (e.g., series-parallel networks), this means that there are no negative externalities due to type differences---the cost of type~$i$ agents may only increase when all others are of the same type.
 We note that this result (along with Prop.~\ref{th:CPk_tight}) strictly generalizes  Theorem~5 from \cite{acemoglu2016informational}, which is attained as a special case for $k=1$ and specific demand types (namely, agents that ignore certain edges); and Theorem~5.7 in \cite{NS15} which is attained as a special case for $k=1$ and symmetric games. It also implies that the Price of Malice~\cite{babaioff2007congestion} is 0 in series-parallel networks (this is since our theorems make no assumption on the behavior of agents of other types).

For larger values of $k$, Theorem~\ref{th:PW_bound} means that regardless of what the other agents are doing, the cost to the type~$i$ agents is never more than their equilibrium cost in a game where all agents are of type $i$, and their number is multiplied by $k$. 
\begin{proof}
\rmr{need to prove a better bound of $C^{i,(1+(k-1)\eps)}$ where $\eps=1-r_i$. The problem is to show that w.l.o.g. all type~$i$ agents use only one edge from $\ol E$. }
Denote $\ol P=P_i(\vec g^*)$ (the set of paths $p\in A_i$ s.t. $g^*_{p,i}>0$).
Assume towards a contradiction that $C^{i,*} > C^{i,(k)}$. This means that $c^i(p,\vec f^*) > C^{i,(k)} = c^i(p,\vec g^*)$ for any path $p\in \ol P$, as $c^i(p,\vec f^*)$ is either $C^{i,*}$ or higher.  For any path $p\in \ol P$, 

$$\sum_{e\in p}c^i_e(f^*_e) = c^i(p,\vec f^*) > c^i(p,\vec g^*) = \sum_{e\in p}c^i_e(g^*_e),$$
thus there is an edge $e=e_p\in p$ s.t. $c^i_e(f^*_e)  > c^i_e(g^*_e)$. Consider the set $E_{\ol P} = \{e_p | p\in \ol P\}$.
 Since $c^i_e$ is monotone, this means $f^*_e > g^*_e$ for every  edge $e\in E_{\ol P}$. 

Consider the weighted directed graph $H=\tup{V, F}$, where $F=\{e\in p | p\in \ol P\}$, and the capacity (weight) of every edge $e\in F$ is $g^*_e$. 
 
 By construction, $E_{\ol P}$ is a cut between $s_i$ and $t_i$ in $H$. Let $\ol E\subseteq E_{\ol P}$ s.t. $\ol E$ is a minimal cut (not necessarily of minimum size or minimum weight), then 
\labeq{cut_flow}
{\sum_{e\in \ol E}f^*_e > \sum_{e\in \ol E}g^*_e \geq k\cdot r.}
The first inequality is since $f^*_e>g^*_e$ holds for every $e \in E_{\ol P}$ and thus for every $e\in \ol E$. The second inequality follows from the min-cut-max-flow theorem, since $\ol E$ is some $s_i-t_i$ cut in $H$,  and the flow between $s_i,t_i$ is exactly $k\cdot r$ in any valid flow of $\GG^{i,(k)}$ (which by definition has a mass of $k\cdot r$ type~$i$ agents). 

Now, as there are only $r$ agents in $\GG=(G,m,(\vec c^i)_{i\leq m},\vec s,\vec t,\vec r)$, then by Eq.~\eqref{eq:cut_flow} and the pigeon-hole principle there must be some agents in profile $\vec f^*$ using strictly more than $k$ edges from the set $\ol E$. Choose some path used by such an agent and denote it by $p'\in A_i$. We define $S= p' \cap \ol E$, thus  by our selection of $p'$, $|S|>k$.

Finally, all edges of $S$ are contained in the minimal cut $\ol E$ between $s_i,t_i$ (thus $S$ is parallel), and $S$ is also contained by the simple path $p'$ (thus $S$ is serial). By definition of serial-parallel width,  $\PW(G|_i) \geq |S| > k$, which is a contradiction to our assumption.  
\end{proof}

\subsection{Lower bound}
We say that a class of cost functions is \emph{nontrivial} if it contains all constant functions, as well as at least one strictly increasing function.
 
Theorem~\ref{th:PW_bound} is tight in a very strong sense: the bound cannot be improved for \emph{any network} and \emph{any nontrivial class} of cost functions. 
\begin{proposition}
\label{th:CPk_tight}
Let any $k\geq 2$,   network $G$ with $\PW(G)=k$, any nontrivial class of cost functions $\calC$, and any $\delta>0$. There is a game with two types $\GG=\tup{G,(\vec c^1,\vec c^2),s,t,\vec r}$, 
such that $c^1_e,c^2_e\in \calC$ for all $e\in E$, and $C^{1,*} > C^{1,(k)}-\delta$.
\end{proposition}
\begin{proof}
We first prove for the $k$-Serial-Parallel network $G_{SP(k)}$.
The demand type of both agents is $(s,t)$, thus all paths from $s$ to $t$ are valid strategies. We define edge costs as follows. Recall that $\calC$ contains all constant functions, and at least one strictly increasing function $c$. For type $i=1$, we set $c_{ab}^i(x)=c(x)$. The cost of all other edges is 0.
For type $j=2$, we set  $c^j_{ab}(x) =  c^j_{ba}(x) =0, c^j_{sa}(x)=c^j_{bt}(x)=c(1)$.
 We set $r_i=\eps$  ($\eps$ will be determined later) and $r_j=1-\eps$.
%

Thus in the unique equilibrium $\vec g^*$ of game $\GG^{i,(k)}$, the agents ($k$ in total) split evenly over all $k$ short paths, and $C^{i,(k)}=c^i_{ab}(\frac{k}{k}) =  c^i_{ab}(1)=c(1)$. 

In the unique equilibrium $\vec f^*$ of game $\GG$, all type $j$ agents take the long path $p=(s,b_1,a_2,b_2,\ldots,b_{k-1},a_k,t)$ since $c^j(p,\vec f^*)=0$. The type~$i$ agents split evenly over all $k$ short paths , and thus each experiences a cost of 
$C^{i,*}=c^i_{ab}(r_j+\frac{r_i}{k}) = c(r_j+\frac{r_i}{k}) >  c(r_j).$  

Let $\eps>0$ such that $c(1-\eps)>c(1)-\delta$. Since $c$ is continuous and strictly increasing such $\eps$ must exist.
Finally,\\ 
$C^{i,*} > c(r_j) = c(1-\eps) > c(1)-\delta = c(1) =C^{i,(k)}-\delta$, as required.

\medskip
Now, consider an arbitrary network $G$ with $\PW(G)\geq k$. 
Consider some serial-parallel set $S$ of size $k$ and the path $p$ that contains it, and $S$ will play the role of the $ab$ edges above. Specifically, 
For type~$i$ agents we set $c^i_e\equiv c^i_{ab}\equiv c$ for any $e\in S$, and $c^i_e\equiv c\equiv 0$ for all other edges. For type~$j$, we set $c^j_e\equiv 0$ for all $e\in p$, and $c^j_e\equiv c(1)$ for all other edges. It is easy to verify that $C^{i,*},C^{i,(k)}$ are the same as for $G_{SP(k)}$. 
%
\end{proof}

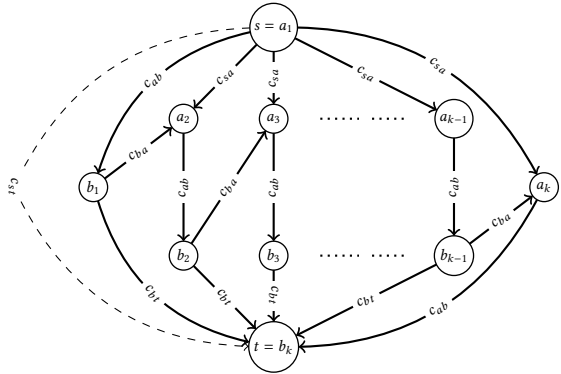
\begin{figure}

\centering
\begin{tikzpicture}[scale=0.6,transform shape]

  \Vertex[L=$b_1$,x=0,y=3]{b1}
  \Vertex[L=$a_2$,x=2,y=4.5]{a2}
  \Vertex[L=$b_2$,x=2,y=1.5]{b2}
	\Vertex[L=$a_3$,x=4,y=4.5]{a3}
	\Vertex[L=$b_3$,x=4,y=1.5]{b3}
  \Vertex[L=$a_{k-1}$,x=8,y=4.5]{ak1}
	\Vertex[L=$b_{k-1}$,x=8,y=1.5]{bk1}
	\Vertex[L=$a_{k}$,x=10,y=3]{ak}
	\Vertex[L=${s=a_1}$,x=4,y=6.5]{s}
  \Vertex[L=${t=b_k}$,x=4,y=-0.5]{t}
	  \tikzstyle{VertexStyle}=[rotate=-90,above]
	\Vertex[L=$c_{st}$,x=-2,y=3]{dummy}
		\Vertex[L=$$,x=4.8,y=4.5]{a4d}
	\Vertex[L=$$,x=4.8,y=1.5]{b4d}
	\Vertex[L=$$,x=7,y=4.5]{ak2d}
	\Vertex[L=$$,x=7,y=1.5]{bk2d}
  \tikzstyle{VertexStyle}=[fill=black!20!white]
\tikzstyle{VertexStyle}=[shape=coordinate]
  \tikzstyle{LabelStyle}=[fill=white,sloped]
  \tikzstyle{EdgeStyle}=[->]

	\Edge[label=$c_{sa}$](s)(a2)
	\Edge[label=$c_{sa}$](s)(a3)
  \Edge[label=$c_{sa}$](s)(ak1)
 
	\Edge[label=$c_{ab}$](a2)(b2)
	\Edge[label=$c_{ab}$](a3)(b3)
	\Edge[label=$c_{ab}$](ak1)(bk1)
 
	\Edge[label=$c_{bt}$](b2)(t)
	\Edge[label=$c_{bt}$](b3)(t)
	\Edge[label=$c_{bt}$](bk1)(t)
	
  \Edge[label=$c_{ba}$](b1)(a2)
	\Edge[label=$c_{ba}$](b2)(a3)
	\Edge[label=$c_{ba}$](bk1)(ak)
	\tikzstyle{EdgeStyle}=[-,dotted]
	\Edge[label=$$](a4d)(ak2d)
	\Edge[label=$$](b4d)(bk2d)
	\tikzstyle{EdgeStyle}=[bend right]
	\tikzstyle{EdgeStyle}=[->,bend right]
	  \Edge[label=$c_{ab}$](s)(b1)
		 \Edge[label=$c_{bt}$](b1)(t)	
		
			\tikzstyle{EdgeStyle}=[dashed,thin,bend right]
 \Edge(s)(dummy)
\tikzstyle{EdgeStyle}=[->,dashed,thin,bend right]
		 \Edge(dummy)(t)	
		
	
	\tikzstyle{EdgeStyle}=[->,bend left]
	 \Edge[label=$c_{sa}$](s)(ak)
	\Edge[label=$c_{ab}$](ak)(t)

\end{tikzpicture}

%
\caption{\label{fig:CPk}The solid edges compose the network $G_{SP(k)}$.  The edge labels are used in Prop. \ref{th:CPk_tight} and \ref{th:CPk_smooth_tight}, where edges with the same label (e.g. $c_{ab}$) have the same cost function. The dashed $(s,t)$ edge is not part of $G_{SP(k)}$ and is used only in Prop.~\ref{th:CPk_smooth_tight}.} 
\end{figure}


\section{Biased Price of Anarchy}
\label{sec:bias}


Given an NRG $\GG=\tup{G,m,\vec c,\vec s,\vec t,\vec r}$ and  modified cost functions $\hat c_e^i$ for every type $i\leq m$ and edge $e\in E$, we obtain a  PNRG $\hat \GG=\tup{G,m,(\hatv c^i)_{i\leq m},\vec s,\vec t,\vec r}$. 
We assume that agents act based on their modified cost functions, irrespective of whether this is a rational behavior or not. 
We refer to $\hat \GG$ as the \emph{biased game}, where every agent of type $i$ experiences a cost of $\hat c^i_e$ rather than $c_e$. 
Both games $\GG$ and $\hat \GG$ have a role in our model, and we often denote the overall setting  as $\GBC=\stup{\GG,\hat \GG}$. 
The way we interpret $\GBC$ is that agents play the biased game $\hat \GG$ (and thus, it is the equilibria of $\hat \GG$ that matter), whereas their true costs are according to 
game $\GG$. We say that $\GBC$ is \emph{uniform} if all agents in $\hat \GG$ have the same cost type (same bias).



\paragraph{Biased Price of Anarchy}

We measure the price of anarchy in a game with biased costs by comparing the equilibria of $\hat \GG$ (denoted by $\hatv f^*$) to the optimum of $\GG$. Formally:

\labeq{b_PoA}{ \BPoA(\GBC)=\BPoA(\GG,\hat \GG) = \sup_{\hatv f^*\in EQ(\hat \GG)} \frac{\mathit{SC}(\GG,\hatv f^*)}{\mathit{OPT}(\GG)}.}

In the uniform bias case where $\hat{\vec c}^i=\hat{\vec c}$ for all $i$, the game $\hat \GG$ is just another NRG. 
%
 Chen et al.~\shortcite{chen2014altruism} referred to the BPoA (when applied to altruism) as the \emph{Robust PoA}. 
The \emph{Price of Risk Aversion}~\cite{NS15} and the \emph{Deviation Ratio}~\cite{kleer2016impact} are similar concepts to the BPoA, except they compare $\hatv f^*$ to the equilibrium of the unbiased game $\GG$. 

\rmr{change to altruism?}
A simple example of a biased cost is induced by {\em pessimism}, which is one way to model risk-aversion~\cite{MP15}.  Suppose that whenever faced with some expected load
$f_e$ on edge $e$, an agent 
takes a safety margin by playing as if the actual load is $\alpha \cdot
f_e$ (for some fixed private parameter $\alpha\geq 1$).
Such an agent will play \emph{as if} every cost function $c_e$ is replaced with a new cost function $\hat c^\alpha_e$, where $\hat c^\alpha_e(x) = c_e(\alpha x)$ (see Fig.~\ref{fig:Braess}). We denote the (uniform) game where all agents play according to $(\hat c^\alpha_e)_{e\in E}$ by $\hat \GG^\alpha$. 

\begin{example}\label{ex:BPoA}
In the optimal flow of the objective game $\GG$ in Fig.~\ref{fig:Braess}, the amount of agents taking the long path is $\sim0.155$, and  thus $\mathit{OPT}(\GG)\cong 1.23$. The equilibrium $\vec f^*$ of $\GG$  is suboptimal (as all agents take the long path $s-a-b-t$), and $\PoA(\GG) \cong \frac{2}{1.23} = 1.626$. In contrast, in the equilibrium $\hatv f^*$ of  $\hat \GG^\alpha$ for $\alpha=3$ (Fig.~\ref{sfig:Braess_r}), agents divide equally among the two short paths, which leads to $\BPoA(\GG,\hat \GG^\alpha) \cong \frac{1.25}{1.23} = 1.016$.
\end{example}

Interestingly, the bias of the agents on Example~\ref{ex:BPoA} somewhat mitigates the increase in social cost that is incurred by rational behavior. 
\subsection{Smoothness for Biased Costs}\label{sec:bias_smooth}

Our goal is to provide bounds on the biased Price of Anarchy for a given game with biased costs $\stup{\GG,\hat \GG}$. That each of $c_e$ and $\hat c^i_e$ are smooth is insufficient to provide such a bound. 

\begin{example}
Consider a Pigou game with pessimistic agents $\GBC_\alpha =\stup{\GG_\alpha,\hat \GG_\alpha^\alpha}$ where $\GG_\alpha=\GG_P(2/\alpha)$ and thus $\hat \GG_\alpha^\alpha=\GG_P(2)$. For any $\alpha$ the equilibrium of $\hat \GG_\alpha^\alpha$ (and thus of $\GBC_\alpha$) is the same: 1/2 of all agents use each resource. However as $\alpha$ increases, the optimal flow of $\GG_\alpha$ shifts more agents to the  resource with variable cost, and the optimal social cost decreases to $0$. Thus the gap between the equilibrium cost and the optimal cost (the BPoA) goes to infinity with $\alpha$ even though both of $\GG_\alpha,\hat \GG^\alpha_\alpha$ are affine.
\end{example}


It is crucial, then, to extend the definition of smoothness to games with biased costs in a way that takes into account both $c$ and $\hat c$. This technique  has been applied before for specific modified costs, for example nonatomic games with restricted tolls~\cite{bonifaci2011efficiency} and atomic games with altruistic agents~\cite{chen2011robust}. We provide a general extension.
%
\begin{definition}
\label{def:p_smooth}
Let $\lambda\geq 0, \mu<1$. The function $c$ is $(\lambda,\mu)$-\emph{biased-smooth} w.r.t. biased cost function $\hat c$, if for any $x,x'\in \mathbb R_+$,
\labeq{b_smooth}{c(x)x + \hat c(x)(x'-x) \leq \lambda c(x')x' + \mu c(x)x.}
\end{definition}
\medskip

It is instructive to check the familiar case where there is no bias. 
Indeed, if $\hat c=c$, and $c$ is $(\lambda,\mu)$-smooth, then 
$$c(x)x + \hat c(x)(x'-x) = c(x)x + c(x)(x'-x) =c(x)x' \leq \lambda c(x')x' + \mu c(x)x,$$
and Eq.~\eqref{eq:b_smooth} collapses to ``standard'' smoothness (Eq.~\eqref{eq:smooth}).

Recall that the PoA of a $(\lambda,\mu)$-smooth game is bounded by $\frac{\lambda}{1-\mu}$. 
This bound extends to games with biased costs that are biased-smooth and when all agents have the \emph{same bias}. This was explicitly shown for specific biases, but we write down the general formulation for completeness.
\begin{theorem}[Bonifaci et al.~\cite{bonifaci2011efficiency}]
\label{th:b_smooth}
Consider a uniform game with biased costs $\GBC=\tup{\GG,\hat\GG}$ where every cost function $c_e$ is $(\hat \lambda,\hat \mu)$-biased smooth w.r.t. biased cost function $\hat c_e$. Let $\hatv f^*\in EQ(\hat \GG)$, and $\vec f$ any valid flow. Then
$\mathit{SC}(\GG,\hatv f^*) \leq \frac{\hat \lambda}{1-\hat \mu} \mathit{SC}(\GG,\vec f)$.\\
In particular, $\BPoA(\GG,\hat \GG)\leq \frac{\hat \lambda}{1-\hat \mu}$.
\end{theorem}
\begin{proof}
The proof extends the standard proof of PoA bounds for nonatomic congestion games via smoothness arguments. 
In any equilibrium $\hatv f^*$,  the \emph{variational inequality} $\sum_e \hat c_e(\hat f^*_e)\hat f^*_e \leq \sum_e \hat c_e(\hat f^*_e) f_e$ holds 
(see \cite{correa2008geometric,R09}).
Thus, 
\begin{align*}
\mathit{SC}(\GG,\hatv f^*) & \leq \mathit{SC}(\GG,\hatv f^*)  +  \sum_e \hat c_e(\hat f^*_e) f_e - \sum_e \hat c_e(\hat f^*_e)\hat f^*_e \\
&\leq \sum_e [\hat \lambda c_e(f_e)f_e + \hat \mu c_e(\hat f^*_e)\hat f^*_e] 
=\hat  \lambda \mathit{SC}(\GG,\vec f) + \hat\mu \mathit{SC}(\GG,\hatv f^*). 
\end{align*}
We get the bound by reorganizing the terms. The only part that differs from the standard smoothness is the first inequality.
\end{proof}



\paragraph{An alternative derivation of optimal tolls}
We can also check that the extension provides the right
result in regard to modified costs $\tilde c(x) = c(x) + c'(x)x$ (where $c'(x)=\frac{\partial c(x)}{\partial x}$) that represent
optimal tolls and should lead to optimal play~\cite{Beckmann56}. Let's confirm this result via a biased-smoothness argument.
\begin{observation}
\label{lemma:opt_toll}
If we set modified cost $\tilde c(x) = c(x) + c'(x)x$, then cost function $c$ is $(1,0)$-biased smooth w.r.t. to $\tilde c$. 
\end{observation} 
This follows immediately from the convexity of $xc(x)$, and affirms via Theorem~\ref{th:b_smooth} that $\BPoA(\GG,\tilde \GG)=1$ for any game. 
For example, adding optimal tolls to the game in Fig.~\ref{sfig:Braess} will set $\tilde c_{sa}(x) = \tilde c_{bt}(x) = 3x^2$, and in the only equilibrium, $SC(\GG,\vec f^*) = 1.23 = OPT(\GG)$. 
%
%
%




\subsection{Bounds for Specific Biases}
\label{sec:h_bounds}
Theorem~\ref{th:b_smooth} lets us bound the BPoA of a population of agents with \emph{uniform bias}, and in the next sections we will prove theorems that bound the BPoA of populations with \emph{heterogeneous biases}. However, all these theorems require some explicit bounds on the biased smoothness parameters $\hat \lambda$ and $\hat \mu$ in a given scenario (i.e., for the specific bias and specific class of cost functions). 

There are by now many such studied biases. We already introduced pessimism, which depends on a single parameter $\alpha$, and where the biased cost is $\hat c^\alpha_e(x)=c_e(\alpha x)$.  Toll-sensitive agents~\cite{cole2003pricing} have a single parameter $\beta$, and their biased cost is $\hat c^{\beta}_e(x) = c_e(x)+\beta x c'_e(x)$ (meaning that marginal tolls are imposed, but agents discount or over-weigh them by a factor of $\beta$).   Altruist agents~\cite{chen2014altruism} have exactly the same biased cost function as tolls, except that now $\beta$ should be interpreted as how much they care about hurting others.  In the Mean-Var risk-aversion model~\cite{nikolova2014mean,NS15}, $c_e(x)$ is a distribution over costs, and the biased cost is $\hat c_e^{\gamma}(x) = \mathbb E[c_e(x)] +\gamma VAR[c_e(x)]$, i.e., $\gamma$ represents the sensitivity of agents to variance.  We assume that $VAR[c_e(x)]\leq \tau$ for some constant $\tau$. \emph{Small error} means that $c$ and $\hat c$ are within a small multiplicative factor of $1+\delta$ from one another (see Sec.~\ref{sec:Psi} for details). 

\begin{figure*}[t]
\begin{center}
\begin{small}
\subfloat[Maximal values of $\hat \lambda,\hat\mu$ under various biases]{ 
\begin{tabular}{||l||c|c||c|c||c||c||}
\hline
Bias &\multicolumn{2}{|c||}{Pessimism}& \multicolumn{2}{|c||}{Toll-sensitivity / Altruism} & Risk-Aversion & Small error\\
\hline
 Params. &  $\alpha \le 2$  & $\alpha \ge 2$  &  $\beta \le 1$  & $\beta \ge 1$ & any $\gamma,\tau$ & any $\delta,\hat \delta$\\
							\hline
							\hline
						$\hat \lambda$ & $1$ &  $\frac{\alpha^2}{4\alpha-4}$ & $1$ & $\frac{(\beta+1)^2}{4\beta}$ & $1+\gamma\tau$ & $1+\delta$ \\
						\hline
						$\hat \mu$ & $1+\frac{\alpha^2}{4}-\alpha$ & $0$ & $\frac{(\beta+1)^2}{4}-\beta$ & $0$ & $\frac14$ & $1-\frac{3}{4(1+\hat \delta)}$\\
\hline  
\end{tabular}\quad\quad}\quad\quad\quad
\subfloat[$\BPoA(\GBC^\alpha)$]{ \label{subfig:graph}
%
%
\begin{tikzpicture}[scale=0.75]
\begin{axis}[xlabel=$\alpha$,ylabel=BPoA,width=6cm,height=3.8cm, enlarge x limits = -1]

			\addplot[domain=1:2, blue] {1/((x) - ((x)^2) / 4)};  
		  \addplot[domain=2:7, blue] {(x)^2/(4*(x-1))};
			

		\addplot[domain=1:7, thin,dotted] {1};
\end{axis}
\end{tikzpicture}
}
\end{small}
\vspace{-2mm}
\caption{\label{tab:toll}Biased-smoothness bounds for affine cost functions. For a trivial bias of $\alpha=1,\beta=0$,  $\gamma\tau=0$, or $\delta=\hat \delta=0$, we always get the familiar affine smoothness of $(\hat\lambda=1,\hat\mu=\frac14)$. The bound on Altruism with $\beta\le 1$ is due to Chen et al.~\shortcite{chen2014altruism}. The bound on small error is due to Prop.~\ref{prop:Phi_to_smooth} applied to affine cost functions. The proofs for all other bounds are in Appendix~\ref{sec:biases}.
}
\end{center}
\end{figure*}
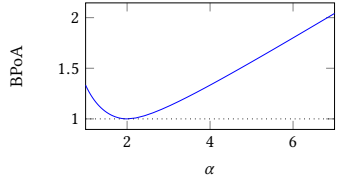

Figure~\ref{tab:toll}  summarizes known biased smoothness bounds for routing games with \emph{affine} cost functions (see Appendix~\ref{sec:biases} for bounds for more general classes of cost functions). Note that from every column in the table we can derive a BPoA bound via Theorem~\ref{th:b_smooth}. For example, for any biased game $\tup{\GG,\hat \GG^\alpha}$ where all cost functions in $\GG$ are affine, and all agents in $\hat \GG$ are pessimistic with parameter $\alpha\leq 2$, we have $\BPoA(\GG,\hat \GG^\alpha)\leq \frac{4}{4\alpha-\alpha^2}$. This bound, which is tight, is illustrated in Fig.~\ref{subfig:graph}. Note that for certain values of $\alpha$, the BPoA is lower than the PoA, i.e., bias steers the society to better outcomes (we also saw this phenomenon in Example~\ref{ex:BPoA}). 

We note that BPoA bounds have also been derived independently using different techniques by Nikolova and Stier-Moses~\shortcite{NS15}, by Meir and Parkes~\shortcite{MP15} and by Kleer at al.~\shortcite{kleer2016impact}.

\section{Structure-dependent Bounds}
\label{sec:diverse}


In this section, we derive a bound, tight up to a constant, on the equilibrium cost for agents participating in a game with heterogeneous biases. This bound depends on their own biased-smoothness parameters, as well as on the structure of the network. In some cases, this also allows us to derive BPoA bounds that depend on the \emph{average bias}.  Most importantly, our bounds in this section do not assume any specific class of cost functions, and work for any function and any bias, as long as biased smoothness holds. 

\subsection{Upper bound}\label{sec:main_statement}
%

 Our primary question is whether we can get a bound on the social cost of any equilibrium of $\GBC$ in terms of
 the smoothness parameters of all types, and the optimal social cost. 

In the next theorem and corollaries, each $c_e$ is $(\lambda,\mu)$-smooth (as per Eq.~\eqref{eq:smooth}), each $\hat c^i_e$ is $(\lambda^i,\mu^i)$-smooth , and $c_e$ is $(\hat{\lambda}^i,\hat{\mu}^i)$-biased-smooth w.r.t. $\hat c^i_e$ (as per Def.~\ref{def:p_smooth}). 
Recall that $\frac{r_i}{r}$ is the fraction of agents of type $i$.

\begin{theorem}[Smoothness theorem]
\label{th:main}
Consider any game with biased costs $\GBC=\tup{\GG,\hat \GG}$ where $\PW(G)\leq k$. 
Let $\hatv f^*$ be an equilibrium of $\hat \GG$. Then for any type $i$, 
\vspace{-2mm}
	  $$SC_i(\GG,\hatv f^*)  \leq \frac{r_i}{r} \frac{\lambda \lambda^i \hat  \lambda^i }{(1-\sqrt \mu)^2 (1-\mu^i) (1-\hat \mu^i)}\frac{1}{k} OPT(\GG^{i,(k)}).$$
%
\end{theorem}
In simple words, the theorem says that for every agent type, the equilibrium cost to all $r_i$ agents of this type may increase (compared to the optimal cost) by a factor that only depends on the smoothness of cost functions, their own bias, and the structure of the network, but not on the biases of other types. This result is bi-criteria, as we compare to the optimum of a game with higher demand. Since costs are increasing functions, $\frac{1}{k}OPT(\GG^{i,(k)})$ is \emph{increasing} in $k$.

\rmr{add proof?}
\begin{lemma}
\label{lemma:hat_c_bounds}
For every $x>0$, and every type $i$, it holds that:

(a) $\hat c^i(x) \geq (1-\hat \mu^i)c(x)$;\footnote{\label{fn:hat_mu}For some biases (e.g. tolls), we have a tighter inequality $\hat c^i(x)\geq c(x)$ from the definition. In such cases we can eliminate the factor $1-\hat \mu^i$ in Theorem~\ref{th:main} and all following corollaries.} 
and (b) $\hat c^i(x) \leq \frac{\lambda \hat  \lambda^i }{(1-\sqrt \mu)^2}  c(x)$.
\end{lemma}
\begin{proof}
For (a), define quantity $x'=0$, so that by $(\hat \lambda^i,\hat \mu^i)$-biased-smoothness,
$$x(c(x) -\hat c^i(x)) = c(x)x + \hat c^i(x)(x'-x)\leq \hat \lambda^i c(x')x' + \hat \mu^i c(x)x = \hat \mu^i c(x)x.$$
Hence $c(x) -\hat c^i(x) \leq \hat \mu^i c(x)$.
%

For (b), set $q=\sqrt{1/\mu}$ and consider $x'=qx$. 
{\small
\begin{align*}
&c(x)x + \hat c^i(x)(x'-x) \leq  \hat \lambda^i c(x')x' + \hat\mu^i c(x)x &\Rightarrow \\
&c(x)x + \hat c^i(x)(q-1)x \leq   \hat \lambda^i c(qx)qx + \hat\mu^i c(x)x &\Rightarrow \\
&\hat c^i(x) \leq   \frac{q}{q-1} \hat \lambda^i c(qx)  + (\hat\mu^i-1)c(x) 
\leq   \frac{q}{q-1} \hat \lambda^i c(qx)  \tag{$\hat\mu^i\leq 1$}
\end{align*}
}
Since $c$ is $(\lambda,\mu)$-smooth, $c(qx)x \leq \lambda  c(x)x + \mu c(qx)qx$ for every $x>0$, and thus
$c(qx) \leq \frac{\lambda}{1-q\mu}c(x)$.
Finally, 
$$\hat c^i(x) \leq \hat \lambda^i\lambda \frac{q}{(q-1)(1-q\mu)}c(x)=  \frac{\lambda \hat  \lambda^i }{(1-\sqrt \mu)^2}  c(x),$$
as required.
\end{proof}
\begin{proof}[Proof sketch of Theorem~\ref{th:main}]
Let $\hatv g^*$ be an equilibrium of $\hat \GG^{i,(k)}$.  Define $\hat C^{i,*} = \hat c^i(p,\hatv f^*)$ for some used path $p\in P_i(\hatv f^*)$, and $\hat C^{i,(k)} =\hat c^i(p,\hatv g^*)$ for some used path $p\in P_i(\hatv g^*)$. In words, $\hat C^{i,*},\hat C^{i,(k)}$ are the \emph{perceived} costs to agents of type~$i$ in equilibria $\hat f^*$ and $\hat g^*$, respectively.  Note that they are not affected by which used path we choose.

The next steps are: to upper bound the ratio between $rk\cdot\hat C^{i,(k)}=\mathit{SC}(\hat \GG^{i,(k)},\hatv g^*)$ and  $OPT(\GG^{i,(k)})$, using Lemma~\ref{lemma:hat_c_bounds}(b) on every used edge;  and to lower bound $\hat C^{i,*}$ by the \emph{true cost} $c(p,\hatv f^*)$ of path $p$,  using Lemma~\ref{lemma:hat_c_bounds}(a) on all edges of $p$. The most important part is the inequality $\hat C^{i,*}\leq \hat C^{i,(k)}$, which is due to Theorem~\ref{th:PW_bound} applied to the biased game $\hat \GG$. 

Finally, recall that what we need to bound is  social cost for all  type~$i$ agents in the ``real game'' $\GG$, which is $\mathit{SC}_i(\GG,\hatv f^*)$. Since each such type~$i$ agent has a cost of exactly $c(p,\hatv f^*)$, we now chain all the previous inequalities to get the desired upper bound. 
\end{proof}

\rmr{The dependence on the structural parameter $k$ is tight, see Sec.~\ref{sec:negative}. Other parameters in the bound can sometimes be relaxed. For example we know that w.l.o.g. $\lambda^i=1$ for any class of functions~\cite{correa2004selfish}, $\lambda=1$ for at least some classes of functions\rmr{ (see Appendix.~\ref{apx:lambda_mu})},
and $(1-\hat \mu^i)$ may be omitted for some biases (see Footnote~\ref{fn:hat_mu}).   
}

\subsection{Implications}
Theorem~\ref{th:main} has a number of useful corollaries. Some use the fact that for series-parallel networks $k=1$ (Prop.~\ref{th:PW_SP}).  
\begin{corollary}\label{cor:main_all}
Consider any  game with biased costs $\GBC=\tup{\GG,\hat \GG}$ where $\PW(G)\leq k$. 
Let $\hatv f^*$ be an equilibrium of $\GBC$. For any type $i$:
\begin{enumerate}
	\item if $\GG$ is affine, $SC_i(\GG,\hatv f^*)  \leq 4\frac{r_i}{r}  \frac{\lambda^i \hat  \lambda^i }{(1-\mu^i) (1-\hat \mu^i)}k \cdot OPT(\GG^{i})$;
	\item if type $i$ is unbiased,  $SC_i(\GG,\hatv f^*)  \leq \frac{r_i}{r} \frac{\lambda}{1-\mu} \frac{1}{k}OPT(\GG^{i,(k)})$.
\end{enumerate}
Moreover, if $\GG$ is symmetric, then:
\begin{enumerate}[resume]
	\item  $SC(\GG,\hatv f^*)  \leq \(\sum_{i\leq m}\frac{r_i}{r} \frac{\lambda \lambda^i \hat  \lambda^i }{(1-\sqrt \mu)^2 (1-\mu^i) (1-\hat \mu^i)}\) \frac{1}{k}OPT(\GG^{(k)})$;
	\item if $\GG$ is affine, $\BPoA(\GBC) 
	\leq O(1)k\sum_{i\leq m}\frac{r_i}{r} \frac{ \hat  \lambda^i }{1-\hat \mu^i}$;
	\item If $G$ is series-parallel, then 
	{\small $\BPoA(\GBC) \leq 
	O(1) \sum_{i\leq m} \frac{r_i}{r} \frac{\hat \lambda^i}{1-\hat \mu^i}$}.
\end{enumerate}
\end{corollary}
These corollaries provide us with more explicit bounds. In particular: (1) for affine cost functions we can get rid of the bi-criteria result since the optimal social cost is linear in the demand (more generally, for polynomials of degree $d$ the factor becomes $O(k^d)$). (2) entails that in a setting where most agents are unbiased and there is only a small fraction of adversarial agents (as in \cite{babaioff2007congestion}),  the damage to the unbiased agents is limited. 
(3)-(5) show that the equilibrium social cost in symmetric games depends on the 
	\emph{average bias} over all agents: each type of biased agents can only increase the social cost by a factor that is proportional to their mass (and also affected by the serial-parallel width of the network). 
Kleer at al.~\shortcite{kleer2016impact} proved  a result similar in spirit to our Corollary~\ref{cor:main_all}(5), but restricted to toll-sensitivity.

\subsection{Lower Bounds}
\label{sec:negative}
Our main theorems both rely on the restricted structure of the underlying network. Prop.~\ref{th:CPk_tight} already shows the necessity of this restriction when some agents behave adversarially.

 We next show that the neither smoothness parameters $(\hat \lambda_i,\hat \mu_i)$ nor the structural parameter $k=\PW(G)$ can be relaxed in Theorem~\ref{th:main} and in its corollaries.

\begin{proposition}\label{th:CPk_smooth_tight}
Let $k\geq 2$, any smoothness parameters $(\hat \lambda^i,\hat \mu^i)$, and any $r_i< \frac12$. There is an affine symmetric game with biases $\GBC=\stup{\GG,\hat \GG}$ with $\PW(G)=k$, such that in the unique equilibrium $\hatv f^*$ of $\hat \GG$, the cost for all type~$i$ agents increases by a factor of $\Omega(1)\frac{\hat \lambda^i}{1-\hat \mu^i}k$ compared to $OPT(\GG)$.
\end{proposition}
\begin{proof}[Proof sketch]
The proof uses the $k$-Serial-Parallel network $G_{SP(k)}$ with one additional edge $e_{st}$ (see Fig.~\ref{fig:CPk}). We define the real costs as follows: $c_{sa}=c_{bt}=c_{ba}\equiv 0, c_{st}\equiv \frac{q}{2},  c_{ab}(x)=x$, where $q=\frac{\hat \lambda^i}{1-\hat \mu^i}$. Note all edges of the form $(a_i,b_i)$ have the same cost $c_{ab}$ and likewise for the other edges. 
  In the flow of $\GG$ where all agents split evenly, the cost is $C^o=\frac{1}{k}$. In the biased game, $\hat c_{ab}^i(x) = qx$ (thus $\frac{\hat \lambda^i}{1-\hat \mu^i}\leq \frac{q}{1-0}=q$), and we modify all other agents so they prefer the long path that intersects all short paths. Thus in equilibrium $\hatv f^*$ all type $i$ agents use edge $e_{st}$ for a cost of $\frac{q}{2}\geq \frac{\hat \lambda^i}{2(1-\hat \mu^i)}\geq \frac{\hat \lambda^i}{2(1-\hat \mu^i)}k\cdot C^o$.
\end{proof}
Since we only used affine cost functions, the bounds in both Theorem~\ref{th:main} and Cor.~\ref{cor:main_all}(1) are tight up to a constant. Also, since the network used in the proof is embedded in  $G_{SP(k+1)}$, and thus embedded in any acyclic network with $\PW(G)>k$ (due to \cite{meir17embedded}), the example in Prop.~\ref{th:CPk_smooth_tight} can be constructed for \emph{any acyclic network} $G$ with $\PW(G)\geq k+1$.


%
\section{Structure-independent Bounds}\label{sec:Psi}

In this section, we leverage known PoA bounds in routing games with player-specific costs~\cite{gairing2011routing} to obtain BPoA bounds in biased games with heterogeneous agents. These are bounds that do not depend on the network structure, but only on the cost functions and agents' biases. 

The bounds in this section are typically worse than those in Section~\ref{sec:diverse}, as they depend on the worst bias rather than the average bias, and this dependency is polynomial rather than linear. Yet, these bounds hold regardless of network structure. 


For two  cost functions $c,\hat c$ and $r>0$, denote by $\Delta(c,\hat c,r) = \sup_{x\in [0,r]}\frac{c(x)}{\hat c(x)}$. Let $\Psi(\GG)= \max_{i,j\leq m}\max_{e\in E}\Delta(c^i_e,c^j_e,r)$. 
\begin{theorem}[Gairing et al.~\cite{gairing2011routing}]
\label{th:PoA_PS}
Consider any PNRG $\GG=\tup{G,(\vec c^i)_{i\leq m},\vec r}$  where  $\vec c^i$ are polynomials of degree $d$ for all $i$. 
Then $\PoA(\GG)  \leq (d+1)\cdot \Psi(\GG)^d$.\footnote{The tight bound ( Thm.~4.10 in Gairing et al.~\shortcite{gairing2011routing}) is better by a factor of up to $(d+1)$, however we are mainly interested in the fact that it is independent of the network structure, and so we use a simplified form.}
\end{theorem}

Intuitively, the PoA is low if for each edge $e$ and any $x$, all functions $(c_e^i(x))_{i\leq m}$ attain similar values. 
We can use this PoA bound to prove a  similar BPoA bound that is independent of the network structure.
  Given a game with biased costs $\GBC=\tup{\GG,\hat \GG}$, let:
	
	\begin{itemize}
		\item $\ol \Phi(\GBC)=\max_{i\leq m}\max_{e\in E}\Delta(\hat c^i_e,c_e,r)$;
		\item $\ul \Phi(\GBC)=\max_{i\leq m}\max_{e\in E}\Delta(c_e, \hat c^i_e,r)$; 
		\item and $\Psi(\GBC)=\Psi(\hat \GG)$.
	\end{itemize}
We first observe that $\Delta$ is tightly related to biased-smoothness, as biased-smoothness bounds entail bounds on $\Delta$, and vice versa. 

%


\def\eu{\mathrm{e}}
\begin{proposition}\label{prop:Phi_to_smooth}
Consider $c,\hat c$ such that $c$ is $(\lambda,\mu)$-smooth,  $\Delta(c,\hat c,r)\leq 1+\delta$, and $\Delta(\hat c, c, r)\leq 1+\hat \delta$ for some $\delta,\hat\delta\geq 0$ and any $r>0$. Then $c$ is $\((1+\delta)\lambda,\frac{\mu+ \hat \delta}{1+\hat\delta}\)$-biased-smooth w.r.t. $\hat c$. 
\end{proposition}

\begin{proposition}\label{prop:smooth_to_Phi}  
Consider $c,\hat c$ such that $c$ is $(\lambda,\mu)$-smooth, and  $(\hat \lambda,\hat \mu)$-biased smooth w.r.t. $\hat c$. Then for any $r>0$ 
$\Delta(c,\hat c,r) \leq \frac{1}{1-\hat \mu}$; $\Delta(\hat c,c,r) \leq \frac{\lambda}{(1-\sqrt \mu)^2}\hat \lambda $. 
Also, if $c$ is a polynomial of degree at most $d$, $\Delta(\hat c,c,r)\leq (d+1)\cdot \eu \cdot \hat \lambda$, where $\eu$ is the natural logarithm base. 
 \end{proposition} 
Proposition~\ref{prop:smooth_to_Phi} follows from Lemma~\ref{lemma:hat_c_bounds} and from the smoothness of polynomial functions. 
Next, we derive a BPoA bound. 
Due to tightness results on the PoA~\cite{gairing2011routing}, we cannot hope to significantly improve the bound in this approach. 
	
\begin{theorem}\label{th:BPoA_PS}
Consider any game  $\GBC=\tup{\GG,\hat \GG}$  where all of $\hat c_e^i$ are polynomials of degree at most $d$. 
Then 
$\BPoA(\GBC)  \leq 
(d+1) (\ul \Phi(\GBC) \ol \Phi(\GBC))^{d+1}.$
\end{theorem}
\begin{proof}[Proof sketch]
Consider an equilibrium flow $\hatv f^*\in EQ(\hat \GG)$. We use the definitions of $\ul \Phi,\ol \Phi$ to show that
 $SC(\GG,\hatv f^*) \leq \ul \Phi(\GBC) \ol \Phi(\GBC) \PoA(\hat \GG)$. We then apply the bound $\PoA(\hat \GG) \leq (d+1)\cdot \Psi(\GG)^d$ from Theorem~\ref{th:PoA_PS}, and show that $\Psi(\hat \GG)=\Psi(\GBC)\leq \ul \Phi(\GBC)\ol \Phi(\GBC)$.
\end{proof}

Finally, we can combine Proposition~\ref{prop:smooth_to_Phi} and Theorem~\ref{th:BPoA_PS} to obtain a  BPoA bound that depends only on the smoothness parameters of the ``worst'' type:
\begin{corollary}\label{th:BPoA_const}
Consider any game with biased costs $\GBC=\stup{\GG,\hat \GG}$  where  for all $i\leq m$ and $e\in E$: (a) $c_e,\hat c_e^i$ are polynomials  of degree at most $d$; and (b) $c_e$ is $(\hat \lambda,\hat \mu)$-biased smooth w.r.t. $\hat c^i_e$. \\
Then 
$\BPoA(\GBC)  \leq 
 (d+1)^{d+2}\eu^{d+1}\(\frac{\hat \lambda}{1-\hat \mu}\)^{d+1}.$
\end{corollary}

\section{Conclusion and Future Work}



We have considered strategic settings in which participants are
making routing decisions based on individually perceived costs. From the perspective of a system analyst who considers the objective costs, the agents' behavior deviates from optimal self-interested play.  Whether these  deviations
 come from a cognitive limitation, subjective preferences, a behavioral
bias, or external influence such as tolls,
it is important to understand how the equilibria of the game are
affected.

Our work is the first to
provide (Biased) PoA bounds for populations with heterogeneous
arbitrary biases, and the first to consider heterogeneous biases in
general networks.  Our main results (Sec.~\ref{sec:ex} and \ref{sec:diverse}) bound the
equilibrium cost of each agent type separately, based only on their
own biases and on the structure of the network. When considering the
entire population, this entails a bi-criteria bound on the social cost
(in some cases can be written as a Biased PoA bound) that depends on the \emph{average}
bias in the population.  
These results can help estimate the worst case effect when the exact structure of the population is  not known in advance: one only has to know the smoothness bounds of each type and a bound on the quantity of each type.

Since in nonatomic routing games all mixed and correlated
equilibria have the same social cost anyway~\cite{blum2006routing}, smoothness does not add more ``robustness,'' and is thus perhaps considered less interesting in such games. 
 Our results show that smoothness is
useful for a \emph{different kind of robustness}, namely to heterogeneous
biases. 

An additional result (Sec.~\ref{sec:Psi}) relies on the PoA of the heterogeneous ``wrong'' game,
and while it does not require the network to have a particular
structure, it provides bounds stated in terms of the \emph{worst} bias in the
population, and only applies for polynomial cost functions.
We can think of these results as bounding the negative externalities of each type \emph{on all others}, whereas the results in Sections~\ref{sec:ex} and \ref{sec:diverse} bound the externalities that all others may inflict \emph{on agents of type~$i$}. 

%

We hope that our results and techniques will inspire further
progress in understanding both the effect of network topology and the effect of bounded rationality on equilibria.

Some challenges are to derive BPoA bounds that only depend on the average bias (i.e., not on the network structure); and bounds on the negative externality that become negligible for a small amount of adversarial agents.  

Biased smoothness can also be applied to obtain robust PoA bounds  in \emph{atomic} congestion games and other normal form games (i.e. for mixed and correlated equilibria), by extending the standard smoothness definition as in \cite{chen2014altruism}. An important open question is whether atomic games with heterogeneous biases can be similarly analyzed, by showing biased-smoothness independently for each type of agent.

 %
%

\bibliographystyle{ACM-Reference-Format}

\bibliography{smoothness}
\appendix

\section{Structure-dependent Bounds}\label{sec:structure}

\begin{lemma}
\label{lemma:two_bounds}
	 $\mathit{SC}(\hat \GG^{i,(k)},\hatv g^*) \leq  \frac{\lambda \lambda^i \hat  \lambda^i }{(1-\sqrt \mu)^2 (1-\mu^i) } OPT(\GG^{i,(k)})$.
\end{lemma}
\begin{proof}
Since by definition $\hat \GG^{i}$ (and thus also $\hat \GG^{i,(k)}$) is $(\lambda^i,\mu^i)$-smooth,  for any flow $\vec f$ of $\hat \GG^{i,(k)}$, 

\labeq{two_bounds}
{\mathit{SC}(\hat \GG^{i,(k)},\hatv g^*)\leq  \frac{\lambda^i}{1-\mu^i} \mathit{SC}( \hat \GG^{i,(k)},\vec f).
}
We denote the optimal flow in $\GG^{i,(k)}$ by $\vec g^o= \vec f^o(\GG^{i,(k)})$. 

\begin{align*}
\mathit{SC}(\hat \GG^{i,(k)},\hatv g^*)&\leq  \frac{\lambda^i}{1-\mu^i} \mathit{SC}( \hat \GG^{i,(k)},\vec g^o) \tag{By~\eqref{eq:two_bounds} with $\vec f=\vec g^o$} 
= \frac{\lambda^i}{1-\mu^i} \sum_{e\in E} g^o_e \hat c^i_e(g^o_e) \\
&\leq  \frac{\lambda^i}{1-\mu^i} \sum_{e\in E} g^o_e \(\frac{\lambda \hat  \lambda^i }{(1-\sqrt \mu)^2} c_e(g^o_e)\) \tag{from Lemma~\ref{lemma:hat_c_bounds}(b)}\\
&= \frac{\lambda \lambda^i \hat  \lambda^i }{(1-\sqrt \mu)^2 (1-\mu^i)} \sum_{e\in E}  g^o_e c_e(g^o_e)
= \frac{\lambda \lambda^i \hat  \lambda^i }{(1-\sqrt \mu)^2 (1-\mu^i)}OPT(\GG^{i,(k)}),
\end{align*}
as required.
%
%
%
\end{proof}

\medskip
\begin{rtheorem}{th:main}
Consider any game with biased costs $\GBC=\tup{\GG,\hat \GG}$ where $\PW(G)\leq k$. 
Let $\hatv f^*$ be an equilibrium of $\hat \GG$. Then for any type $i$, 
\vspace{-2mm}
	  $$SC_i(\GG,\hatv f^*)  \leq \frac{r_i}{r} \frac{\lambda \lambda^i \hat  \lambda^i }{(1-\sqrt \mu)^2 (1-\mu^i) (1-\hat \mu^i)}\frac{1}{k} OPT(\GG^{i,(k)}).$$
\end{rtheorem}

\begin{proof}
Similarly to the definitions before Theorem~\ref{th:PW_bound}, we define $\hatv f^*$ as an equilibrium of $\hat \GG$, $\hat C^{i,*} = \hat c^i(p,\hatv f^*)$ for some used path $p\in P_i(\hatv f^*)$, and $\hat C^{i,(k)} =\hat c^i(p,\hatv g^*)$ for some used path $p\in P_i(\hatv g^*)$. Note that $\hat C^{i,*},\hat C^{i,(k)}$ are not effected by which used path we choose.

%
%

By Theorem~\ref{th:PW_bound} applied to games $\hat \GG,\hat \GG^{i,(k)}$,
{\small
$$r \cdot \hat C^{i,*}  \leq r \cdot \hat C^{i,(k)} =  \frac{1}{k}(r \cdot k)  \hat C^{i,(k)} = \frac{1}{k}(\sum_{p\in A} \hat g^*_p) \hat C^{i,(k)} 
 = \frac{1}{k}\!\!\sum_{p\in P^i(\hatv g^*)}\!\!\! \hat g^*_p  \hat c^i(p,\hatv g^*) = \frac{1}{k}\mathit{SC}(\hat \GG^{i,(k)},\hatv g^*).$$}

On the other hand, for any used path $p\in P^i(\hatv f^*)$, by Lemma~\ref{lemma:hat_c_bounds}(a)
$$\hat C^{i,*}=  \hat c^i(p,\hatv f^*) = \sum_{e\in p} \hat c^i_e(\hat f^*_e) \geq  \sum_{e\in p}(1-\hat\mu^i)  c_e(\hat f^*_e) = (1-\hat\mu^i) c(p,\hatv f^*).$$

Combining the above bounds, we get an upper bound on the total cost for all  type~$i$ agents:
{\small
$$SC_i(\GG,\hatv f^*) = \sum_{p\in A}\hat f^*_{i,p} c(p,\hatv f^*) \leq \sum_{p\in A}\hat f^*_{i,p} \frac{\hat C^{i,*}}{1-\hat \mu^i}  
= \frac{r_i\hat C^{i,*}}{1-\hat \mu^i}  =\frac{r_i (r\cdot \hat C^{i,*})}{r (1-\hat \mu^i)} 
\leq \frac{r_i}{r}\frac{1}{(1-\hat \mu^i)} \frac{1}{k}\mathit{SC}(\hat \GG^{i,(k)},\hatv g^*).$$
}
Finally, we get the theorem by plugging in the bound of Lemma~\ref{lemma:two_bounds}.
\end{proof}

\begin{rcorollary}{cor:main_all}
Consider any  game with biased costs $\GBC=\tup{\GG,\hat \GG}$ where $\PW(G)\leq k$. 
Let $\hatv f^*$ be an equilibrium of $\GBC$. For any type $i$:
\begin{enumerate}
	\item if $\GG$ is affine, $SC_i(\GG,\hatv f^*)  \leq 4\frac{r_i}{r}  \frac{\lambda^i \hat  \lambda^i }{(1-\mu^i) (1-\hat \mu^i)}k \cdot OPT(\GG^{i})$;
	\item if type $i$ is unbiased,  $SC_i(\GG,\hatv f^*)  \leq \frac{r_i}{r} \frac{\lambda}{1-\mu} \frac{1}{k}OPT(\GG^{i,(k)})$.
\end{enumerate}
Moreover, if $\GG$ is symmetric, then:
\begin{enumerate}[resume]
	\item  $SC(\GG,\hatv f^*)  \leq \(\sum_{i\leq m}\frac{r_i}{r} \frac{\lambda \lambda^i \hat  \lambda^i }{(1-\sqrt \mu)^2 (1-\mu^i) (1-\hat \mu^i)}\) \frac{1}{k}OPT(\GG^{(k)})$;
	\item if $\GG$ is affine, $\BPoA(\GBC) 
	\leq O(1)k\sum_{i\leq m}\frac{r_i}{r} \frac{ \hat  \lambda^i }{1-\hat \mu^i}$;
	\item If $G$ is series-parallel, then 
	{\small $\BPoA(\GBC) \leq 
	O(1) \sum_{i\leq m} \frac{r_i}{r} \frac{\hat \lambda^i}{1-\hat \mu^i}$}.
\end{enumerate}
\end{rcorollary}

	\begin{proof}
\begin{enumerate}
\item First, for affine functions, $\lambda=1,\mu=\frac14$, thus $\frac{\lambda}{(1-\sqrt \mu)^2} = \frac{1}{1/4}=4$ (by Lemma.~\ref{lemma:lambda_mu} there is no better pair $(\lambda,\mu)$). Second, for any $k\geq 1$ and any flow $\hatv f$ in $\GG$,
$$  SC(\GG,k\cdot \hatv f) = \sum_{e\in E}(k \cdot f_e) c_e(k \cdot f_e) = \sum_{e\in E}(k \cdot f_e) (a_e \cdot k \cdot f_e + b_e) \leq k^2 \sum_{e\in E} f_e (a_e f_e + b_e) = k^2 SC(\GG,\hatv f),$$
thus 
$OPT(\GG^{i,(k)}) = SC(\GG,\vec f^o(\GG^{i,(k)})) \leq SC(\GG,k\cdot \vec f^o(\GG^{i})) \leq k^2 SC(\GG, \vec f^o(\GG^{i}))=k^2 OPT(\GG^{i}).$
We get the corollary by replacing the respective terms in Theorem~\ref{th:main}. 
More generally, for degree-$d$ polynomials, the approximation factor is $O(k^d)$.

\item Strictly speaking, Cor.~\ref{cor:main_all}(2) is not entailed by Theorem~\ref{th:main}. Proof is by following the same steps of the proof of Theorem~\ref{th:main}, except for Lemma~\ref{lemma:hat_c_bounds}, which becomes redundant for unbiased agents (see Footnote~\ref{fn:hat_mu}). Thus $\frac{\lambda \hat \lambda^i}{(1-\sqrt \mu)^2(1-\hat \mu^i)}$ can be omitted from the bound, which leaves us with $\frac{\lambda^i}{1-\mu^i}=\frac{\lambda}{1-\mu}$ since $\hatv c^i=\vec c^i$.  

\item This is since if $\GG$ is symmetric, then  $\GG^{i}=\GG$ and $\GG^{i,(k)}=\GG^{(k)}$ for all $i$ and $k$. Thus
$$SC(\GG,\hatv f^*)=\sum_{i\leq m} SC_i(\GG,\hatv f^*)  \leq \(\sum_{i\leq k}\frac{r_i}{r} \frac{\lambda \lambda^i \hat  \lambda^i }{(1-\sqrt \mu)^2 (1-\mu^i) (1-\hat \mu^i)}\) \frac{1}{k}OPT(\GG^{(k)}).$$

\item From (1), 
\begin{small}
$$SC(\GG,\hatv f^*) = \sum_{i\leq m} SC_i(\GG,\hatv f^*)  \leq \sum_{i\leq m}4\frac{r_i}{r}  \frac{\lambda^i \hat  \lambda^i }{(1-\mu^i) (1-\hat \mu^i)}k \cdot OPT(\GG^{i})
= 4k OPT(\GG)\sum_{i\leq m}\frac{r_i}{r}  \frac{\lambda^i \hat  \lambda^i }{(1-\mu^i) (1-\hat \mu^i)}.$$
\end{small}

\item Follows immediately from (3) with $k=1$.
\end{enumerate}
\end{proof}

\subsection{A single smoothness parameter}
\label{apx:lambda_mu}
For a given class of cost functions, we denote their smoothness parameters by $(\lambda,\mu_\lambda)$, i.e. by fixing $\lambda\geq 1$ (necessary for any class that includes constant functions), and finding the minimal $\mu=\mu_\lambda$ such that all functions in the class are $(\lambda,\mu)$ smooth.
Recall that $\frac{\lambda}{1-\mu_\lambda}$ is always minimized for $\lambda=1$~\cite{correa2004selfish}.
\begin{lemma}\label{lemma:lambda_mu}
For polynomial functions, $\frac{\lambda}{(1-\sqrt\mu_\lambda)^2}$ is minimized for $\lambda=1$. 
\end{lemma}
We suspect that this is true for any function class that includes all constant functions. 

\begin{proof}
Consider any class of polynomial functions of degree at most $d$. It is not hard to show that $\mu_\lambda = \frac{d\lambda}{(\lambda(1+d))^{1+\frac{1}{d}}}=\lambda^{-\frac{1}{d}}\frac{d}{(1+d)^{1+\frac{1}{d}}}$ (similar to the smoothness bounds in \cite{correa2004selfish}).

Thus 
\begin{align*}
\frac{\lambda}{(1-\sqrt\mu)^2}&=\frac{\lambda}{\(1-\sqrt{\lambda^{-\frac{1}{d}}\frac{d}{(1+d)^{1+\frac{1}{d}}}}\)^2}=\frac{\lambda}{\(1-\lambda^{-\frac{1}{2d}}\frac{d}{(1+d)^{\frac{d+1}{2d}}}\)^2}\\
&=\frac{1}{\(\lambda^{-\frac12}-\lambda^{-\frac12}\lambda^{-\frac{1}{2d}}\frac{d}{(1+d)^{\frac{d+1}{2d}}}\)^2}\\
&=\frac{1}{\(\lambda^{-\frac12}-\lambda^{-\frac{d+1}{2d}}\frac{d}{(1+d)^{\frac{d+1}{2d}}}\)^2} = \frac{1}{y(\lambda)^2}\\
\end{align*}

Now, $\frac{\partial \lambda^{-\frac12}}{\partial \lambda} = -\frac12 \lambda^{-\frac32}$, whereas
\begin{align*}
&\frac{\partial\( \lambda^{-\frac{d+1}{2d}}\frac{d}{(1+d)^{\frac{d+1}{2d}}} \)}{\partial \lambda} = -\frac{d+1}{2d}\frac{d}{(1+d)^{\frac{d+1}{2d}}} \lambda^{-\frac{d+3}{2d}}\\
&= -\frac{1}{2(1+d)^{\frac{d+1}{2d}-1}} \lambda^{-\frac{d+3}{2d}}\\
& =-\frac{1}{2(1+d)^{-\frac{d-1}{2d}}} \lambda^{-\frac{d+3}{2d}}\\
&= -\frac12\lambda^{\frac{d+3}{2d}}(1+d)^{\frac{d-1}{2d}} \\
\end{align*}
It thus holds that for the \emph{derivative} of $y(\lambda)$:
\begin{align*}
y'(\lambda) &= -\frac12 \(\lambda^{\frac{d+3}{2d}}(1+d)^{\frac{d-1}{2d}}-\lambda^{-\frac32}\) \leq 0 &\iff \\
& \(\lambda^{\frac{d+3}{2d}}(1+d)^{\frac{d-1}{2d}}-\lambda^{-\frac32}\) \geq 0 &\iff \\
& \lambda^{\frac{d+3}{2d}}(1+d)^{\frac{d-1}{2d}}\geq  \lambda^{-\frac32} &\iff \\
&\lambda^{\frac{d+3}{2d}}\lambda^\frac32 \geq (1+d)^{-\frac{d-1}{2d}}&\iff \\
&\lambda^{\frac{4d+3}{2d}} \geq (1+d)^{-\frac{d-1}{2d}}&\iff\\
&\lambda^{4d+3} \geq (1+d)^{-(d-1)}&\iff\\
&\lambda \geq (1+d)^{-\frac{d-1}{4d+3}} 
\end{align*}
Note that the last right-hand expression is upper-bounded by $1$ for all $d\geq 1$. 
In particular for all $\lambda\geq 1$, $y'(\lambda)$ is non-positive, $y(\lambda)$ is non-increasing, and  $\frac{\lambda}{(1-\sqrt \mu_\lambda)^2}=\frac{1}{y(\lambda)^2}$ is non-decreasing. Hence $\frac{1}{(1-\sqrt \mu_1)^2} \leq \frac{\lambda}{(1-\sqrt \mu_\lambda)^2}$ for all $\lambda\geq 1$. 
%
\end{proof}

\subsection{Lower bounds}
\begin{theorem}[\cite{meir17embedded}]\label{TH:DSP_EMBED}\rmr{fix citation}
Let $G$ be a 2-terminal directed acyclic graph, and let $k\geq 2$. The following conditions coincide:
\begin{enumerate}[topsep=0.5ex,itemsep=0.1ex,labelindent=1em]
  \item $G$ is a directed series-parallel graph;
	\item The directed Braess graph $G_B$ is not d-embedded in $G$;
	\item $\PW(G)= 1$;
	\end{enumerate}
\end{theorem}
\begin{proposition}\label{prop:lower_Braess}
Consider any graph such that $\PW(G)>1$. For any $\eps>0$ and any $M'>0$,  there is a symmetric game with biases  $\GBC=(\GG,\hat \GG)$ and equilibria $\vec f^*\in EQ(\GG), \hatv f^*\in EQ(\hat \GG)$, such that:
\begin{itemize}
	\item Only a fraction $r_2=\eps$ of the agents in $\GBC$ are biased;
	\item For the unbiased type~$1$, $SC_1(\GG,\hat{\vec f}^*)\geq M' \cdot SC_1(\GG,{\vec f}^*)$;
	\item In particular, $\BPoA(\GBC) > (1-\eps)M'$.
\end{itemize}
\end{proposition}
\begin{figure}
\centering
\subfloat[$c(x)=\hat c^1(x)$]{ 
\begin{tikzpicture}[scale=0.6,transform shape]

  \Vertex[x=0,y=3]{a}
  \Vertex[x=4,y=3]{b}
  \Vertex[x=2,y=5]{s}
  \Vertex[x=2,y=1]{t}
  \tikzstyle{VertexStyle}=[fill=black!20!white]
  
  \tikzstyle{LabelStyle}=[fill=white,sloped]
  \tikzstyle{EdgeStyle}=[->]
  \Edge[label=$x$](s)(a)
  \Edge[label=$0$](s)(b)
  \Edge[label=$0$](a)(t)
  \Edge[label=$x$](b)(t)
  \Edge[label=$M$](a)(b)
\end{tikzpicture}
}
~~~~~~~
\subfloat[$\hat c^2(x)$]{\label{sfig:G2}
\begin{tikzpicture}[scale=0.6,transform shape]

  \Vertex[x=0,y=3]{a}
  \Vertex[x=4,y=3]{b}
  \Vertex[x=2,y=5]{s}
  \Vertex[x=2,y=1]{t}
  \tikzstyle{VertexStyle}=[fill=black!20!white]
  
  \tikzstyle{LabelStyle}=[fill=white,sloped]
  \tikzstyle{EdgeStyle}=[->]
  \Edge[label=$0$](s)(a)
  \Edge[label=$M$](s)(b)
  \Edge[label=$M$](a)(t)
  \Edge[label=$0$](b)(t)
  \Edge[label=$0$](a)(b)
\end{tikzpicture}
}
%
\caption{\label{fig:network} All agents need to select an $s-t$  path. Some fraction of type~2 agents adopt the modified cost function $\hat c^2$, where
$M\gg 1$ is some large constant.}
\end{figure}
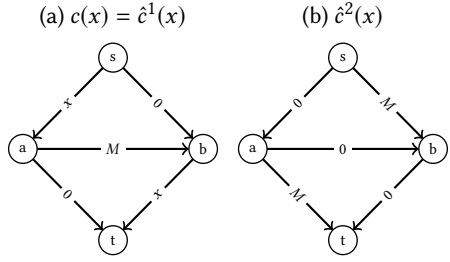
\begin{proof}
 By theorem~\ref{TH:DSP_EMBED} it is sufficient to construct a bad example on the Braess graph to prove the proposition.


Indeed, consider the network in Figure~\ref{fig:network}, where the value $M$ is some large number that will be set later on.
Suppose the total mass is one unit, and that all agents need to
flow from $s$ to $t$. In the game $\GG$, costs are as in 
Figure~\ref{fig:network}~(a). Thus in the equilibrium $\vec f^*$, the agents will split evenly among the paths $s-a-t$ and
$s-b-t$, so that each agent incurs the same cost of $(2\cdot 1/2)^M = 1^M=1$, and thus $SC_1(\vec f^*,\GG)=r_1=1-\eps$. Now, denote the unbiased agents as type~1, and
suppose that some fraction $r_2=\eps>0$ of agents is of a type~2,
with biased costs $\hat c^2$. See Figure~\ref{fig:network}~(b). 

In the equilibrium $\hatv f^*$, all type~2 
agents will select the path $s-a-b-t$, whereas all type~1 agents will
split evenly. Thus there will be $\hat f^*_e = \frac12+\frac{\eps}{2} >f^*_e$ agents on each edge
$s-a$ and $b-t$. 

We now compute the cost for the type $1$ agents in $\hat {\vec f}^*$. Set $M> \frac{M'}{\eps}$. 
For every such agent taking the path $p=(s-a-t)$, the cost is 
$$c(p,\hat {\vec f}^*) = c_{ua}(\hat f^*_{ua}) = c_{ua}(2(\frac12+\frac{\eps}{2}))^M = (1+\eps)^M > M'.$$ 
Thus the total cost for type~1 agents is $SC_1(\GG,\hat{\vec f}^*) > (1-\eps) M' = M' SC_1(\GG,\vec f^*)$.

Finally, since $\vec f^*$ is a valid flow in $\GG$, 
$$\BPoA(\GBC) \geq \frac{SC(\hatv f^*,\GG)}{SC(\vec f^*,\GG)} \geq \frac{SC_1(\hatv f^*,\GG)}{SC(\vec f^*,\GG)} > \frac{(1-\eps)M'}{1} = (1-\eps)M',$$
 as required. 
\end{proof}

\begin{rproposition}{th:CPk_smooth_tight}
Let any $k\geq$, any $q>2$, and any $r_i< \frac12$. There is a symmetric game with biases $\GBC=\stup{\GG,\hat \GG}$, and a flow $\hatv f^*$ s.t.:
\begin{enumerate}
	\item $\PW(\GG)=k$;
	\item in $\hat \GG$ there are $r_i$ agents of type $i$, and $\sum_{j\leq m} r_j=1$;
	\item $\stup{\GG,\hat \GG^i}$ is $(\hat \lambda^i,\hat \mu^i)$-biased smooth, where $\frac{\hat \lambda^i}{1-\hat \mu^i}=q$;
	\item $\hatv f^*$ is the unique equilibrium of $\hat \GG$; 
	\item $SC_i(\GG,\hatv f^*) \geq \Omega(1)\frac{r_i}{r k}\frac{\hat \lambda^i}{1-\hat \mu^i}OPT(\GG^{i,(k)}) =\Omega(1)\frac{r_i}{r}\frac{\hat \lambda^i}{1-\hat \mu^i}k OPT(\GG)$.
\end{enumerate}
\end{rproposition}

\begin{proof}
Consider the $k$-Cross-Parallel graph $G_{CP(k)}$  (see Fig.~\ref{fig:CPk}), with one additional edge $e_{st}$. We define the real costs as follows: $c_{sa}=c_{bt}=c_{ba}\equiv 0, c_{st}\equiv \frac{q}{2},  c_{ab}(x)=x$. There is a mass of $r=1$ agents, and all paths from $s$ to $t$ are allowed (the game is symmetric). In the equilibrium flow of $\GG^{i,(k)}$  the mass $rk=k$ splits evenly among all $k$ parallel paths (excluding $e_{st}$),  and the cost for each agent is $c_{ab}(1) = 1$. In particular, $OPT(\GG^{i,(k)}) = k$. Similarly, in the flow of $\GG$ where all agents split evenly, $OPT(\GG) = r\cdot c_{ab}(\frac{1}{k}) = \frac{1}{k}$.

We now define the modified costs. For type~$i$, $\hat c_{ab}^i(x) = qx$, whereas all other costs remain unchanged. For all other agents (denote as type $j$), we set $\hat c^j_{st} = \hat c^j_{sa} = \hat c^j_{bt} \equiv 10qk$ and $\hat c^j_{ab}=\hat c^j_{ba} \equiv 0$. 
We argue that $(\GG,\hat \GG^i)$ is $(q,0)$-biased smooth: We know (see Table~\ref{tab:toll})) that any affine function $c(x)$ is $(\frac{q^2}{4(q-1)},0)$-biased smooth w.r.t. $\hat c(x)=c(qx)$ for any $q\geq 1$.   Note that  $\frac{q^2}{4(q-1)} \leq q$ for all $q>\frac43$, and thus $\frac{\hat \lambda^i}{1-\hat \mu^i}\leq \frac{q}{1-0}=q$.  

In any equilibrium of $\hat \GG$, all type $j$ agents follow the long path through all vertices $(s,b_1,a_2,b_2,\ldots,b_{k-1},a_k,t)$. Hence for any path that includes an $a-b$ edge, the cost for $i$ is at least $\hat c^i_{ab}(r_j) = q \cdot r_j   >\frac{q}{2}$. Thus in the only equilibrium $\hatv f^*$, all type~$i$ agents will select the direct path $e_{st}$ and experience a cost of $\frac{q}{2}$. 

Putting everything together,
$$SC_i(\GG,\hatv f^*) =  r_i \frac{q}{2} = \frac12 \frac{r_i}{rk} q k = \Omega(1)\frac{r_i}{r k}\frac{\hat \lambda^i}{1-\hat \mu^i}OPT(\GG^{i,(k)}) = \Omega(1)\frac{r_i}{r}\frac{\hat \lambda^i}{1-\hat \mu^i}k OPT(\GG),$$
as required.
\end{proof}
Since we only used affine cost functions, the bounds in both Theorem~\ref{th:main} and Cor.~\ref{cor:main_all}(1) are tight up to a constant. Also, since the graph used in the proof is a minor of  $G_{CP(k+1)}$, and thus by Theorem~\ref{TH:DSP_EMBED} a minor of any graph with $\PW(G)>k$, the example in Prop.~\ref{th:CPk_smooth_tight} can be constructed for \emph{any graph} $G$ with $\PW(G)>k$.

\section{Structure-independent Bounds}\label{sec:no_structure}
\rmr{the bound is not tight}
\begin{rproposition}{prop:Phi_to_smooth}
Consider $c,\hat c$ such that $c$ is $(\lambda,\mu)$-smooth,  $\Delta(c,\hat c,r)\leq 1+\delta$, and $\Delta(\hat c, c, r)\leq 1+\hat \delta$ for some $\delta,\hat\delta\geq 0$. Then $c$ is $\((1+\delta)\lambda,\frac{\mu+ \hat \delta}{1+\hat\delta}\)$-biased-smooth w.r.t. $\hat c$ (in the range $[0,r]$). 
\end{rproposition}
\begin{proof}
Consider some $x,x'\in [0,r]$. If $x'\geq x$, then
\begin{align*}
c(x)x + \hat c(x)&(x'-x) \leq c(x)x + (1+\delta)c(x)(x'-x) \\
&= c(x)x +(1+\delta)c(x)x' -c(x)x - \delta c(x)x = (1+\delta)c(x)x'  - \delta c(x)x\\
&\leq (1+\delta)(\lambda c(x')x' +\mu c(x)x) - \delta c(x)x = (1+\delta)\lambda c(x')x' +(\mu + \delta (\mu - 1))c(x)x\\
& \leq (1+\delta)\lambda c(x')x' +\mu c(x)x 
\leq (1+\delta)\lambda c(x')x' +\frac{\mu+ \hat \delta}{1+\hat\delta} c(x)x \tag{$\mu<1$}\\
\end{align*}
If $x>x'$, then 
\begin{align*}
c(x)x + \hat c(x)&(x'-x) = c(x)x - \hat c(x)(x-x') \leq c(x)x- \frac{1}{1+\hat\delta}c(x)(x-x') \\
&= (1- \frac{1}{1+\hat\delta})c(x)x  + \frac{1}{1+\hat\delta}c(x)x'\\
&=\frac{\hat \delta}{1+\hat\delta}c(x)x  + \frac{1}{1+\hat\delta}c(x)x' \leq \frac{\hat \delta}{1+\hat\delta}c(x)x  + \frac{1}{1+\hat\delta}(\lambda c(x')x' + \mu c(x)x)\\
&=\frac{\lambda}{1+\hat\delta} c(x')x' + \frac{\hat \delta+\mu}{1+\hat\delta}   c(x)x\\
&\leq \lambda c(x')x' + \frac{\mu+\hat \delta}{1+\hat\delta} c(x)x
\leq (1+\delta)\lambda c(x')x' + \frac{\mu+ \hat \delta}{1+\hat\delta} c(x)x
\end{align*}
We therefore get that for any $x,x'\geq 0$, $c(x)x + \hat c(x)(x'-x)\leq (1+\delta)\lambda c(x')x' + (\mu + \hat\delta(1-\mu)) c(x)x$, meaning that $c$ is $\((1+\delta)\lambda,\mu + \hat\delta(1-\mu)\)$-biased smooth w.r.t $\hat c$. 
\end{proof}

\begin{rproposition}{prop:smooth_to_Phi}
Let $r>0$. 
Suppose that $c$ is $(\lambda,\mu)$-smooth, and  $(\hat \lambda,\hat \mu)$-biased smooth w.r.t. $\hat c$. Then 
$\Delta(c,\hat c,r) \leq \frac{1}{1-\hat \mu}$; $\Delta(\hat c,c,r) \leq \frac{\lambda}{(1-\sqrt \mu)^2}\hat \lambda$.
 Also, for polynomials of degree at most $d$, $\Delta(\hat c,c,r)\leq (d+1)\eu \hat \lambda$, where $\eu$ is the natural logarithm base. 
 \end{rproposition} 
\begin{proof}
By definition, $\Delta(c,\hat c,r)  \leq \sup_{x>0}\frac{c(x)}{\hat c(x)}$, which is upper bounded by $\frac{1}{1-\hat \mu}$ due to Lemma~\ref{lemma:hat_c_bounds}(a). Similarly, $\Delta(c,\hat c,r)  \leq \sup_{x>0}\frac{\hat c(x)}{c(x)} \leq \frac{\lambda}{(1-\sqrt \mu)^2}\hat \lambda$ due to Lemma~\ref{lemma:hat_c_bounds}(b).

Due to biased-smoothness, for all $x,x'\geq 0$, $c(x) +\hat c(x)(x'-x)\leq \hat \lambda c(x')x' + \hat \mu c(x)x$. 
Thus for polynomials of degree $d$, 

%
%
\begin{align*}
&\hat c(x)(x'-x) \leq \hat \lambda c(x')x' + (\hat \mu -1)c(x)x &\Rightarrow \\
&\hat c(x)(x'-x) \leq \hat \lambda c(x')x'  &\Rightarrow \\
&\hat c(x)\eps x \leq \hat \lambda c((1+\eps)x)(1+\eps)x  &\Rightarrow \tag{For $x'=(1+\eps)x$}\\
&\hat c(x)\eps \leq \hat \lambda (1+\eps)^d c(x)(1+\eps)   &\Rightarrow \tag{$c$ is degree $d$ polynomial}\\
&\hat c(x) \leq \hat \lambda \frac{1}{\eps}(1+\eps)^{d+1} c(x)  &\\
&\leq \hat \lambda \frac{1}{1/(d+1)}\(1+\frac{1}{d+1}\)^{d+1} c(x)  & \tag{For $\eps = \frac{1}{d+1}$}\\
&\leq (d+1)\cdot \eu \cdot \hat \lambda c(x),& 
\end{align*}
as required.
\end{proof}
\begin{rtheorem}{th:BPoA_PS}
Consider any game with biased costs $\GBC=\tup{\GG,\hat \GG}$  where  $\hatv c^i$ are polynomial functions for all $i$. 
Then 
$\BPoA(\GBC)  \leq \ul \Phi(\GBC) \ol \Phi(\GBC) \PoA(\hat \GG) \leq (d+1) (\ul \Phi(\GBC) \ol \Phi(\GBC))^{d+1}.$
\end{rtheorem}
\begin{proof}
Consider games $\GG$ and $\hat \GG$, both over the network $G=(V,E)$. For any flow $\vec f$, we have 
\begin{align*}
SC(\GG,\vec f) &= \sum_{e\in E}f_e c_e(f_e) =  \sum_{e\in E}\sum_{i\leq m}f_{e,i} c_e(f_e) \leq \sum_{e\in E}\sum_{i\leq m}f_{e,i} \ul \Phi(\GBC) \hat c^i_e(f_e) \\
&= \ul\Phi(\GBC) \sum_{i\leq m}\sum_{e\in E}f_{e,i} \hat c^i_e(f_e) = \ul \Phi(\GBC) \sum_{i\leq m} SC_i(\hat \GG,\vec f)
\leq \ul\Phi(\GBC) SC(\hat \GG,\vec f).
\end{align*}
Similarly, $SC(\hat \GG,\vec f) \leq \ol \Phi(\GBC) SC(\GG,\vec f)$ for every flow $\vec f$.  

For $\hatv f^*\in EQ(\hat \GG), \hatv f^o = \hatv f^o(\GG)$, we have
\begin{align*}
SC(\GG,\hatv f^*)&\leq \ul\Phi(\GBC) SC(\hat \GG,\hatv f^*) \leq \ul \Phi(\GBC) \PoA(\hat \GG) SC(\hat \GG,\hatv f^o) \\
&\leq \ul \Phi(\GBC) \PoA(\hat \GG) \ol \Phi(\GBC) SC(\GG,\hatv f^o)\\
& = \ul \Phi(\GBC) \ol \Phi(\GBC) \PoA(\hat \GG),
\end{align*}
Due to  Theorem~\ref{th:PoA_PS},
 $\BPoA(\GBC) \ol \Phi(\GBC) \ul \Phi(\GBC) \PoA(\hat \GG)   \leq (d+1)\ul \Phi(\GBC) \Psi(\GBC)^d \ol \Phi(\GBC)$, which is a constant independent of the network $G$.

Finally, 
\begin{align*}
\Psi(\GBC) &= \Psi(\hat \GG)= \max_{i,j\leq m}\Delta(\hatv c^i,\hatv c^j,r) = \max_{i,j\leq m}\max_{e\in E}\sup_{x\in [0,r]}\frac{\hat c^i_e(x)}{\hat c^j_e(x)}\\
&=\max_{i,j\leq m}\max_{e\in E}\sup_{x\in [0,r]}\frac{\hat c^i_e(x)}{c_e(x)}\frac{c_e(x)}{\hat c^j_e(x)}\\
&\leq \max_{i\leq m}\max_{e\in E}\sup_{x\in [0,r]}\frac{\hat c^i_e(x)}{c_e(x)}\max_{i\leq m}\max_{e\in E}\sup_{x\in [0,r]}\frac{c_e(x)}{\hat c^j_e(x)}\\
&=\max_{i\leq m}\Delta(\hatv c^i,\vec c,r)\max_{i\leq m}\Delta(\vec c,\hatv c^j,r) = \ol \Phi(\GBC) \ul \Phi(\GBC),
\end{align*}
which entails the theorem.
\end{proof}

\begin{rcorollary}{th:BPoA_const}
Consider any game with biased costs $\GBC=\tup{\GG,\hat \GG}$  where  for all $i\leq m$ and $e\in E$: (a) $c_e,\hat c_e^i$ are polynomials  of degree at most $d$; and (b) $c_e$ is $(\hat \lambda,\hat \mu)$-biased smooth w.r.t. $\hat c^i_e$. 
Then 
$\BPoA(\GBC)  \leq (d+1)\eu\frac{\hat \lambda}{1-\hat \mu}\PoA(\hat \GG)  \leq (d+1)^{d+2}\eu^{d+1}\(\frac{\hat \lambda}{1-\hat \mu}\)^{d+1}.$
\end{rcorollary}
\begin{proof}
By Prop.~\ref{prop:smooth_to_Phi},
\begin{align*} 
&\ul \Phi(\GBC) = \max_{i\leq m}\max_{e\in E}\Delta(c,\hat c^i,r) = \max_{i\leq m}\max_{e\in E}\sup_{x\in [0,r]}\frac{c_e(x)}{\hat c^i_e(x)} \leq\max_{i\leq m}\frac{1}{1-\hat \mu^i}\leq \frac{1}{1-\hat \mu};\\
&\ol \Phi(\GBC) = \max_{i\leq m}\max_{e\in E}\Delta(\hat c^i, c,r) = \max_{i\leq m}\max_{e\in E}\sup_{x\in [0,r]}\frac{\hat c^i_e(x)}{c_e(x)} \leq\max_{i\leq m} ((d+1)\cdot \eu \cdot \hat \lambda^i)= (d+1)\cdot \eu \cdot \hat \lambda.
\end{align*}
By combining the above bounds with Theorem~\ref{th:BPoA_PS}, the proof is complete.
\end{proof}


\section{Specific Biases}\label{sec:biases}

The examples of pessimism, risk aversion, and small error reflect cognitive biases that cause agents to optimize the ``wrong cost.''  In the examples of toll-sensitivity and altruism, the agents may be aware of their own cost functions, but these are still different from the one that the analyst cares about.


\begin{figure}
\centering
\subfloat[$c(x)$]{ \label{sfig:Braess_real}
\begin{tikzpicture}[scale=0.75,transform shape]

  \Vertex[x=0,y=3]{a}
  \Vertex[x=4,y=3]{b}
  \Vertex[x=2,y=5]{s}
  \Vertex[x=2,y=1]{t}
  \tikzstyle{VertexStyle}=[fill=black!20!white]
  
  \tikzstyle{LabelStyle}=[fill=white,sloped]
  \tikzstyle{EdgeStyle}=[->]
  \Edge[label=$x^2$](s)(a)
  \Edge[label=$1$](s)(b)
  \Edge[label=$1$](a)(t)
  \Edge[label=$x^2$](b)(t)
  \Edge[label=$0$](a)(b)
\end{tikzpicture}
}
~~~~~~~
\subfloat[$\hat c^\alpha(x) = c(3x)$]{ \label{sfig:Braess_alpha}
\begin{tikzpicture}[scale=0.75,transform shape]

  \Vertex[x=0,y=3]{a}
  \Vertex[x=4,y=3]{b}
  \Vertex[x=2,y=5]{s}
  \Vertex[x=2,y=1]{t}
  \tikzstyle{VertexStyle}=[fill=black!20!white]
  
  \tikzstyle{LabelStyle}=[fill=white,sloped]
  \tikzstyle{EdgeStyle}=[->]
  \Edge[label=$9x^2$](s)(a)
  \Edge[label=$1$](s)(b)
  \Edge[label=$1$](a)(t)
  \Edge[label=$9x^2$](b)(t)
  \Edge[label=$0$](a)(b)
\end{tikzpicture}
}
\\
\subfloat[$\hat c^\beta(x) = c^*(x)$]{\label{sfig:Braess_beta}
\begin{tikzpicture}[scale=0.75,transform shape]

  \Vertex[x=0,y=3]{a}
  \Vertex[x=4,y=3]{b}
  \Vertex[x=2,y=5]{s}
  \Vertex[x=2,y=1]{t}
  \tikzstyle{VertexStyle}=[fill=black!20!white]
  
  \tikzstyle{LabelStyle}=[fill=white,sloped]
  \tikzstyle{EdgeStyle}=[->]
  \Edge[label=$3x^2$](s)(a)
  \Edge[label=$1$](s)(b)
  \Edge[label=$1$](a)(t)
  \Edge[label=$3x^2$](b)(t)
  \Edge[label=$0$](a)(b)

\end{tikzpicture}
}
~~~
\subfloat[$\hat c^{\gamma,\tau}(x) = c(x)+\frac12x$]{\label{sfig:Braess_gamma}
\begin{tikzpicture}[scale=0.75,transform shape]

  \Vertex[x=0,y=3]{a}
  \Vertex[x=4,y=3]{b}
  \Vertex[x=2,y=5]{s}
  \Vertex[x=2,y=1]{t}
  \tikzstyle{VertexStyle}=[fill=black!20!white]
  
  \tikzstyle{LabelStyle}=[fill=white,sloped]
  \tikzstyle{EdgeStyle}=[->]
  \Edge[label=$x^2+\frac12x$](s)(a)
  \Edge[label=$1+ \frac12x$](s)(b)
  \Edge[label=$1+ \frac12x$](a)(t)
  \Edge[label=$x^2+\frac12x$](b)(t)
  \Edge[label=$\frac12x$](a)(b)

\end{tikzpicture}
}
\caption{Fig.~\ref{sfig:Braess_real} illustrates a game $\GG$ that is a quadratic variation of the Braess paradox~\cite{frank1981braess}. The other subfigures present  three biased versions of $\GG$. For example, the game $\hat \GG_b$ is  played by pessimistic agents, with parameter $\alpha=3$. Hence  the edges with fixed costs do not change, but the biased cost on the edge $a-b$ for example is $\hat c_{s-a}(x)= c_{s-a}(3x)=(3x)^2=9x^2$.  \label{fig:example_Braess}
}
\end{figure}
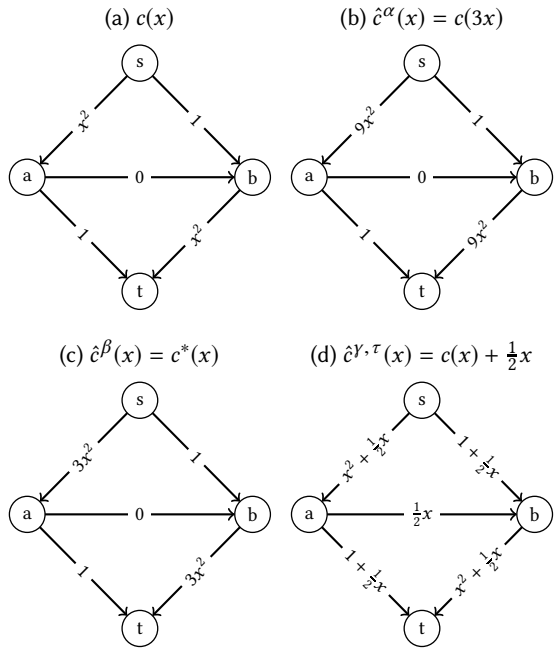

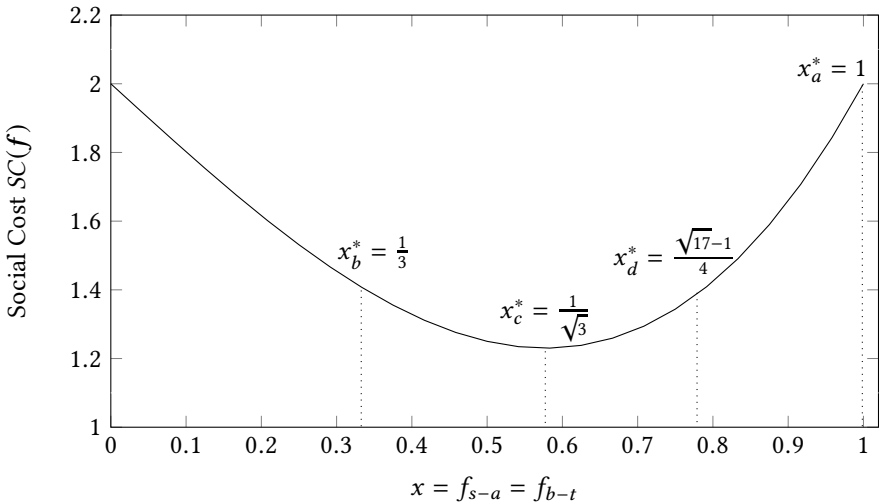
\begin{figure}
\centering

\begin{tikzpicture}[scale=1]
\begin{axis}[xlabel={$x=f_{s-a}=f_{b-t}$},ylabel={$\textrm{Social Cost }SC(\vec f)$} ,width=11.8cm,height=7cm, enlarge x limits = -1]
    \addplot[domain=0:1] {2*x^3 -2*x+2}; 
		\addplot[domain=0:1.02,dotted,white]{2.1};
		\addplot[domain=0:1.02,dotted,white]{1.1};
\end{axis}
\draw[dotted,thin] (10,4.5) -- (10,-0); 
\node[color=black] at (9.6,4.7) {$x^*_a=1$};
\draw[dotted,thin] (3.33,1.8) -- (3.33,-0); 
\node[color=black] at (3.5,2.3) {$x^*_b=\frac{1}{3}$};
\draw[dotted,thin] (5.78,1) -- (5.78,-0); 
\node[color=black] at (5.78,1.43) {$x^*_c=\frac{1}{\sqrt{3\mock}}$};
\draw[dotted,thin] (7.8,1.77) -- (7.8,-0); 
\node[color=black] at (7.5,2.32) {$x^*_d=\frac{\sqrt{17\mock}-1}{4}$};
\end{tikzpicture}
\caption{ The  social cost in game $\GG$ from Fig.~\ref{fig:example_Braess} as a function of the traffic on the edge $s-a$ (assuming symmetric flow). The equilibria of all four games  are presented. \label{fig:example_eq}}
\end{figure}
%


\def\calG{\mathcal{G}}
\subsection{Uncertainty and Risk Aversion}
Following Nikolova and Stier-Moses~\shortcite{nikolova2014mean,NS15}, we consider an arbitrary cost function $c(x)$ and
an arbitrary distribution $\eps(x)$ with mean 0 and bounded variance $\mathrm{var}(\eps(x))\leq
\tau c(x)$ for all $x>0$.  Denote $v_\eps(x) = \frac{\mathrm{var}(\eps(x))}{c(x)}\in [0,\tau]$, and suppose that $c(x)$ is $(1,\mu)$-smooth for some $\mu<1$.
For an agent with \emph{risk aversion} $\gamma$, define the biased cost as 
\begin{align}\hat c^{\gamma,\eps}(x) = c(x) + \gamma v_\eps(x)c(x),\tag{Risk-averse cost}\end{align}
 (meaning the agent prefers paths that are longer in expectation, but with lower variance).

Nikolova and Stier-Moses bounded the {\em Price of Risk Aversion} (PRA), which is the ratio between the social welfare in the biased equilibrium and that in the non-biased equilibrium. They focus on a structural parameter $\eta$ of the underlying network (may be as large as $|V|/2$ in the general case). Denote by $\calG(\eta,\mu)$ all games where the network parameter is at most $\eta$, and every cost function is $(1,\mu)$-smooth. We omit parameters if they do not impose any constraints.

 The main result in \cite{NS15} is that the PRA is upper bounded by $1+\tau \gamma \eta$. In particular this leads to a bound of $\BPoA(\GG,\hat \GG^\gamma)\leq \frac{1+\gamma\tau\eta}{1-\mu}$ for any game $\GG\in \calG(\eta,\mu)$, under uniform risk aversion $\gamma$ and a bound $v_\eps(x)\leq \tau$.

We show that $c$ is  $(1+\gamma \tau,\mu)$-biased-smooth w.r.t. $\hat c^{\gamma,\eps}$, which entails a BPoA bound of $\frac{1+\gamma \tau}{1-\mu}$.
\begin{proposition}Let $c$ be any function that is $(1,\mu)$-smooth, and $\eps(x)$ such that $v_\eps(x)\leq \tau$. 
For $\gamma\geq 0$, $c(x)$ is $(1+\gamma \tau,\mu)$-biased-smooth w.r.t. $\hat c^{\gamma,\eps}(x)$. 
\end{proposition} 
\begin{proof}
Suppose first that $x'\leq x$. Then 
\begin{align*}
xc(c) + \hat c^{\gamma,\eps}(x)(x'-x) &= xc(x) + (c(x)+\gamma v_\eps(x)c(x))(x'-x) \leq c(x)x + c(x)(x'-x) \\
& = c(x)x' \leq c(x')x' + \mu c(x)x \leq (1+\gamma\tau)c(x')x' + \mu c(x)x. \tag{from smoothness of $c$}
\end{align*}
Next, suppose that $x'>x$. Then 
\begin{align*}
xc(c) + \hat c^{\gamma,\eps}(x)(x'-x) &= xc(x) + (c(x)+\gamma v_\eps(x)c(x))(x'-x) \leq xc(x) + (c(x)+\gamma \tau c(x))(x'-x) \\
& = xc(x) + (1+\gamma \tau)c(x)(x'-x)  = -\gamma \tau xc(x) + (1+\gamma\tau)c(x)x' \\
& \leq  -\gamma \tau c(x)x + (1+\gamma\tau)(c(x')x' + \mu c(x)x) \\
& = (1+\gamma\tau)c(x')x' + \mu c(x)x + \gamma \tau (\mu-1)c(x)x  \tag{Since $\mu< 1$}\\
& \leq (1+\gamma\tau)c(x')x' + \mu c(x)x, 
\end{align*}
as required.
\end{proof}
\begin{corollary}
For any game $\GG\in \calG(\mu)$, any noise $\eps$ such that $v_\eps(x)\leq \tau$ and any risk aversion parameter $\gamma$, 
$$\BPoA(\GG,\hat \GG^\gamma)\leq \frac{1+\gamma \tau}{1-\mu}.$$
\end{corollary}
Thus in games with homogeneous risk-aversion we can apply either this structure-independent bound or the structure-dependent bound from  Nikolova and Stier-Moses~\shortcite{NS15}. Which bound is better depends on the exact game.

\subsection{Small errors}
The biased cost function $\hat c_e$ can be any function as long as $\frac{\hat c_e(x)}{c_e(x)}\leq 1+\hat \delta$ and $\frac{c_e(x)}{\hat c_e(x)}\leq 1+\delta$ for all $x\leq r$.  

In such games where the objective game $\GG$ is $(\lambda,\mu)$-smooth, we have a BPoA bound of $\frac{(1+\delta)\lambda}{1-\frac{\mu+\hat \delta}{1+\hat \delta}}$ by Prop.~\ref{prop:smooth_to_Phi}.

\medskip
The  previous examples of biased games assumed  players' perceptions that are biased or misguided. The next  example is a case where the objectives of the players and the analyst/designer differ. 

\subsection{Altruism and sensitivity to tolls}

Suppose that a central authority imposes an optimal toll of $xc'_e(x)$ on every edge (where $c'_e(x)$ is the derivative of $c_e$).  
If agents treat the monetary toll and the cost due to delay in the same way, then we get  biased cost function $\hat c_e(x) = c_e(x) + xc'_e(x) = \tilde c_e(x)$, with the effect
that the incentives  are perfectly aligned with those of the society, and the BPoA is 1~\cite{Beckmann56}. We emphasize that the designer in this case is only interested in the cost stemming from congestion, and disregards any monetary transfers. 

However since the basic cost is in terms of delay, and the toll is in terms of money, different agents may have different money value for time, and hence different toll sensitivity~\cite{cole2003pricing,karakostas2004edge,fleischer2004tolls,fotakis2010existence}.  
This suggests a biased cost function 
\begin{align}
\hat c^\beta(x) = c(x) + \beta x c'(x),\tag{toll-sensitive cost}
\end{align}
 (see Fig.~\ref{fig:example_Braess}(c)).  As discussed in Yang and Zheng~\shortcite{yang2008existence} these individual differences may be \emph{unobservable}, and hence cannot be taken into account by the toll mechanism.  Chen et al.~\shortcite{chen2014altruism} used the same biased cost for values $\beta\in [0,1]$, but in their paper the motivation was \emph{altruism} rather than tolls.

In our case, the designer does not know the value of $\beta$ (or its distribution), and does not try to fit the tolls to the present population. Rather, we are interested in the effect of the standard, marginal-cost toll scheme (i.e., optimal for unbiased agents) on the social cost as $\beta$ varies. 

Denote by $\calG(\text{poly}(d))$ the class of games whose cost functions are polynomials of degree at most $d$ with non-negative coefficients (thus $\calG(\text{poly}(1))$ is the class of affine games).

\begin{lemma}\label{lemma:d_use} For any $d>0$, 
$\frac{1}{(1+d)^\frac1d}- \frac{1}{(1+d)^\frac{d+1}{d}}  = \frac{d}{(1+d)^{\frac{d+1}{d}}}$.
\end{lemma}
\begin{proof}
\begin{align*}
 &\frac{1}{(1+d)^\frac1d}- \frac{1}{(1+d)^\frac{d+1}{d}} = \frac{1}{(1+d)^\frac1d} - \frac{1}{(1+d)(1+d)^\frac{1}{d}} 
= \frac{1}{(1+d)^\frac1d}\(1-\frac{1}{1+d}\) \\
& = \frac{1}{(1+d)^\frac1d} \frac{d}{1+d} = \frac{d}{1+d} \frac{1}{(1+d)^\frac1d} = \frac{d}{(1+d)^{\frac{d+1}{d}}}.
\end{align*}
\end{proof}

\begin{proposition}[Independently also proven by Chen et al.~\cite{chen2014altruism}] \label{th:poly_beta_le_1}
For tax sensitivity $\beta\in [0, 1]$, polynomial cost functions of degree $d$ are $\(1,d\(\frac{1+d\beta}{1+d}\)^{\frac{d+1}{d}}-d\beta\)$-biased-smooth w.r.t. $\hat c^\beta$.  Hence for any game $\GG\in \calG(\text{poly}(d))$ and $\beta \in [0, 1]$, we have 
$$\BPoA\(\GG,\beta\)\leq \frac{1}{1+d\beta -d\(\frac{1+d\beta}{1+d}\)^{\frac{d+1}{d}}}.$$
 This is a tight bound.
\end{proposition}
\begin{proof}
We use the inequality~\cite{roughgarden2004bounding} 
\labeq{poly}
{X^d Y \leq d (d+1)^{-\frac{d+1}{d}}X^{d+1} + Y^{d+1}.}

Let $c$ be a polynomial of degree $d^*$, and $x,x'\geq 0$. 
\begin{align*}
c(x)x + & \hat c^\beta(x)(x'-x) = \sum_{d=0}^{d^*}a_d x^d x + (\sum_{d=0}^{d^*}a_d x^d + \beta \sum_{d=0}^{d^*}d a_d x^{d-1} x)(x'-x)\\
&= \sum_{d=0}^{d^*}(a_d x^{d}x' +\beta d a_d x^d x' - \beta d a_d x^{d+1})\\
&= \sum_{d=0}^{d^*}(a_d (1+d \beta) x^{d}x' - \beta d a_d x^{d+1})
\end{align*}
For every $d\leq d^*$, setting $X= x (d\beta+1)^{1/d}; Y=x'$ and reorganizing terms,
\begin{align*}
a_d (1+d \beta) x^{d}x' &= a_d (X^d Y) \leq a_d ( d (d+1)^{-\frac{d+1}{d}}X^{d+1} + Y^{d+1}) \\
& = a_d ( d (d+1)^{-\frac{d+1}{d}} (d\beta+1)^{\frac{d+1}{d}}x^d+ x'^{d+1}) \\
& =  a_d \( d \(\frac{d\beta+1}{d+1}\)^{\frac{d+1}{d}}x^d+ x'^{d+1}\).
\end{align*}
Thus summing over all $d\leq d^*$,
\begin{align*}
c(x)x + & \hat c^\beta(x)(x'-x) = \sum_{d=0}^{d^*}(a_d (1+d \beta) x^{d}x' - \beta d a_d x^{d+1})\\
& \leq \sum_{d=0}^{d^*} a_d \( d \(\frac{d\beta+1}{d+1}\)^{\frac{d+1}{d}}x^d+ x'^{d+1}\) - \beta d a_d x^{d+1} \\
& =  \sum_{d=0}^{d^*} a_d d \( \(\frac{d\beta+1}{d+1}\)^{\frac{d+1}{d}} - \beta\)x^d+ x'^{d+1}\\
& \leq \sum_{d=0}^{d^*} a_d d^* \( \(\frac{d^*\beta+1}{d^*+1}\)^{\frac{d^*+1}{d^*}} - \beta\)x^d+ x'^{d+1}\\
& =  d^* \( \(\frac{d^*\beta+1}{d^*+1}\)^{\frac{d^*+1}{d^*}} - \beta\)\sum_{d=0}^{d^*} a_d x^d+ \sum_{d=0}^{d^*}x'^{d+1}\\
& = d^* \( \(\frac{d^*\beta+1}{d^*+1}\)^{\frac{d^*+1}{d^*}} - \beta \)c(x)x + c(x')x'.
\end{align*}

\rmr{tightness}
For tightness, consider the Pigou example $\GG_P(\frac{1}{1+d\beta},d)$, i.e., where $c_1(x)=1,c_2(x)=\frac{1}{1+d\beta}x^d$. 

The biased cost on edge~2 is $\hat c^\beta_2(x)= c_2(x) + \beta x c'_2(x) = \frac{1+\beta}{1+d\beta}x^{d+1}$.

Let $f^*=f^*_2$ be the equilibrium load on edge~2. We argue that $f^*=1$. Indeed, $\hat c^\beta_2(f^*)  =  \frac{1+\beta}{1+d\beta}1^{d+1} = \frac{1+\beta}{1+d\beta} \leq 1$, thus no agent wants to switch to edge~1.

 This means $\mathit{SC}(\vec f^*) = 1-f^*+a(f^*)^{d+1} = a = \frac{1}{1+d\beta}$.

Now consider the state $f'=\frac{1}{((d+1)a)^{1/d}}=\(\frac{1+d\beta}{1+d}\)^{1/d}$.  We have that 
\begin{align*}
\mathit{SC}(\mathit{OPT})&\leq \mathit{SC}(\vec f') = 1-f'+a
(f')^{d+1} = 1-\(\frac{1+d\beta}{1+d}\)^\frac1d + \frac{1}{1+d\beta}\(\frac{1+d\beta}{1+d}\)^{\frac{d+1}{d}}
\end{align*}

\begin{align*}
 \frac{ \mathit{SC}(\vec f')}{\mathit{SC}(\vec f^*)} 
&= (1+d\beta)\(1-\(\frac{1+d\beta}{1+d}\)^\frac1d + \frac{1}{1+d\beta}\(\frac{1+d\beta}{1+d}\)^{\frac{d+1}{d}}\) \\
& = 1+d\beta - (1+d\beta)\(\frac{1+d\beta}{1+d}\)^\frac1d + \(\frac{1+d\beta}{1+d}\)^{\frac{d+1}{d}} \\
& = 1+d\beta - (1+d\beta)^\frac{d+1}{d}\(\frac{1}{(1+d)^\frac{1}{d}} - \frac{1}{(1+d)^\frac{d+1}{d}}\) \\
& = 1+d\beta - (1+d\beta)^\frac{d+1}{d}\(\frac{d}{(1+d)^{\frac{d+1}{d}}}\) \tag{By Lemma~\ref{lemma:d_use}} \\
& = 1+d\beta - d\(\frac{(1+d\beta)}{(1+d)}\)^{\frac{d+1}{d}}
\end{align*}

Therefore, 
$$\BPoA(\GG_P,\hat G^\beta) \geq \frac{ \mathit{SC}(\vec f^*)}{\mathit{SC}(\vec f')} =  \frac{1}{1+d\beta - d\(\frac{(1+d\beta)}{(1+d)}\)^{\frac{d+1}{d}}},$$
as required.
\end{proof}

\begin{proposition}\label{th:poly_beta_ge_1}
For tax sensitivity $\beta\geq 1$, polynomial cost functions of degree $d$ are $(\frac{(1+d\beta)^{d+1}}{\beta^d (d+1)^{d+1}},0)$-biased-smooth w.r.t. $\hat c^\beta$. 
Hence for any game $\GG\in \calG(\text{poly}(d))$ and $\beta \geq 1$, we have $\BPoA(\GG,\beta)\leq\frac{(1+d\beta)^{d+1}}{\beta^d (d+1)^{d+1}}$. This is a tight bound.
\end{proposition}
\begin{proof}

Let $c$ be a polynomial of degree $d^*$, and $x,x'\geq 0$.  As in the previous proof, 
\begin{align*}
c(x)x + & \hat c^\beta(x)(x'-x) 
= \sum_{d=0}^{d^*}(a_d (1+d \beta) x^{d}x' - \beta d a_d x^{d+1})
\end{align*}
For every $d\leq d^*$, setting $X= x \beta (d+1)^{1/d}; Y=\frac{x' (1+d\beta)}{d+1}$ and reorganizing terms,
\begin{align*}
a_d (1+d \beta) x^{d}x' &= \frac{a_d}{\beta^d}  x^d \beta^d (d+1) \frac{x' (1+d\beta)}{d+1}\\
& = \frac{a_d}{\beta^d} (x \beta (d+1)^{1/d})^d \frac{x' (1+d\beta)}{d+1} = \frac{a_d}{\beta^d} X^d Y,
\end{align*}
 By using inequality~\eqref{eq:poly},
\begin{align*}
a_d (1+d \beta) x^{d}x' &\leq \frac{a_d}{\beta^d} (d (d+1)^{-\frac{d+1}{d}}X^{d+1} + Y^{d+1}) \\
& = \frac{a_d}{\beta^d} ( d (d+1)^{-\frac{d+1}{d}} (x \beta (d+1)^{1/d})^{d+1} + (\frac{x' (1+d\beta)}{d+1})^{d+1})\\
& = a_d \beta d  x^{d+1} + \frac{a_d}{\beta^d} (\frac{x' (1+d\beta)}{d+1})^{d+1})\\
& = a_d \beta d  x^{d+1} + a_d \frac{(1+d\beta)^{d+1}}{\beta^d (d+1)^{d+1}}(x')^{d+1}.\\
\end{align*}
Summing over all $d\leq d^*$,
\begin{align*}
c(x)x + & \hat c^\beta(x)(x'-x)  \leq \sum_{d=0}^{d^*}a_d \frac{(1+d\beta)^{d+1}}{\beta^d (d+1)^{d+1}}(x')^{d+1}\\
&\leq  \sum_{d=0}^{d^*}a_d \frac{(1+d^*\beta)^{d^*+1}}{\beta^{d^*} ({d^*}+1)^{{d^*}+1}}(x')^{d+1} \\
&= \frac{(1+d^*\beta)^{d^*+1}}{\beta^{d^*} ({d^*}+1)^{{d^*}+1}}\sum_{d=0}^{d^*}a_d (x')^{d+1} = \frac{(1+d^*\beta)^{d^*+1}}{\beta^{d^*} ({d^*}+1)^{{d^*}+1}} c(x')x'.
\end{align*}
For tightness, consider the Pigou game $\GG_P(\frac{\beta^d(d+1)^d}{(1+d\beta)^{d+1}},d)$, i.e. $a=\frac{\beta^d(d+1)^d}{(1+d\beta)^{d+1}}$. 
Denote $\gamma = (1+d\beta)^{1/d}$, then $a= (\frac{\beta (d+1)}{\gamma^{(d+1)}})^d$.
In the state $\vec f'$ where $f'_2=1$, we have 
$$SC(\vec f') = 1-f'_2 + a\cdot (f'_2)^d = a.$$
In equilibrium $\vec f^*$, we have 
$$a\cdot (1+\beta d) (f^*_2)^{d} =  \hat c_2(f^*_2) = \hat c_1(f^*_1) = 1,$$
meaning that 
$$f^*_2 = \frac{1}{a^{1/d} (1+\beta d)^{1/d}} = \frac{\gamma^{(d+1)}}{\beta (d+1)}\frac{1}{(1+\beta d)^{1/d}} = \frac{\gamma^{d}}{\beta (d+1)}.$$ 
Finally, 
\begin{align*}
SC(\vec f^*) &= 1-f^*_2+a(f^*_2)^{d+1} = 1-\frac{\gamma^{d}}{\beta (d+1)}+a \frac{\gamma^{d(d+1)}}{\beta^{d+1} (d+1)^{d+1}} \\
& = 1-\frac{\gamma^{d}}{\beta (d+1)}+ (\frac{\beta (d+1)}{\gamma^{(d+1)}})^d \frac{\gamma^{d(d+1)}}{\beta^{d+1} (d+1)^{d+1}}\\
& = 1- \frac{\gamma^{d}}{\beta (d+1)}+ \frac{1}{\beta(d+1)} = 1-\frac{\gamma^d-1}{\beta(d+1)}\\
& = 1 - \frac{d\beta}{\beta(d+1)} = 1-\frac{d}{d+1} = \frac{1}{d+1},
\end{align*}
which gives us $\BPoA(\GG_P,\beta)\geq \frac{SC(\vec f^*)}{SC(\vec f')} = \frac{1}{a(d+1)} = \frac{(1+d\beta)^{d+1}}{\beta^d(d+1)^d(d+1)} = \frac{(1+d\beta)^{d+1}}{\beta^d(d+1)^{d+1}}$.
\end{proof}

\subsection{Pessimism in the Worst-case Cost Model}
\label{sec:risk_WCC}

Suppose now that agents are pessimistic~\cite{MP15}, in the sense that they play according
to a congestion amount that is larger by a factor of $\alpha>1$ than the true
congestion, that is, 
\begin{align}\hat c^{\alpha}(x) = c(\alpha x).\tag{Pessimist cost}\end{align}

%
 For affine cost functions, it holds that 
\labeq{toll_WCC}
{\hat c^\alpha(x) = \alpha ax + b = ax+b + (\alpha-1)ax  = c(x) + (\alpha-1)c'(x)x,}
and pessimism with factor $\alpha\geq 1$ coincides with toll-sensitivity of $\beta = \alpha-1$. 
As a result, we can immediately apply the bounds for
toll-sensitive agents to pessimistic agents with affine cost functions, and these bounds strictly improve those in \cite{MP15}. 

\rmr{add the proposition}
For higher order polynomial $d>1$, the equivalence to toll-sensitivity (Eq.~\eqref{eq:toll_WCC}) no longer applies, and  the BPoA is $\alpha^{\Theta(d)}$ (for homogenous bias $\alpha$).

Another result in \cite{MP15} is regarding agents with heterogeneous pessimism levels. This result only applies for games with  affine costs, and depends on the worst bias in the mix, in the spirit of Theorem~\ref{th:BPoA_const}. 

\rmr{Include bounds from \cite{kleer2016impact}?
add tables from EC'17 submission?}

\section{PoA in Heterogeneous Games}\label{sec:PoA}
\rmr{not to include for EC}
While our focus in this paper is on biased games, some of our insights can be applied to understand PoA in heterogeneous games.
Specifically, we show that PoA in (heterogeneous) PNRGs can be arbitrarily high even when all agents have affine costs.

A natural conjecture would be that $\PoA(\GG)\leq \sum_{i \leq m} \frac{r_i}{r} \frac{\lambda_i}{1-\mu_i}$, or at least $\PoA(\GG)\leq \max{i \leq m} \frac{\lambda_i}{1-\mu_i}$. However, the following simple example shows that this is not the case, even for a simple variation of the Pigou game. 
\begin{proposition} \label{th:bad_PoA_Pigou}
For any $M$, there is a PNRG with affine costs $\GG$ with two parallel links, s.t. $\Psi(\GG)\leq M^2$ and $\PoA(\GG)\geq \Omega(M)$. 
\end{proposition}
\begin{proof}
Suppose we have $r_1 =\frac{1}{2M}$ agents of type~1, with $c^1_a(x)\equiv 1,c^1_b(x)=x$. The other $r_2=1-\frac{1}{2M}$ agents have $c^2_a(x)\equiv  \frac{1}{M^2}$, and $c^2_b(x)= \frac{x}{M^2}$.  Thus in the unique equilibrium $\vec f^*$ (in fact, the dominant strategy equilibrium), all agents use edge $b$, and 
$$SC(\vec f^*) = r_1 \cdot c^1_b(1) + r_2 \cdot c^2_b(1) > r_1 = \frac{1}{2M}.$$ 

On the other hand, consider the flow $\vec f'$, where  type~1 agents use edge $b$, and all  type~2 agents use edge $a$. We have 
$$SC(\vec f') =  r_1 \cdot c^1_b(r_1) + r_2 \cdot \frac{1}{M^2} <  \frac{1}{(2M)^2} + \frac{1}{M^2} < \frac{2}{M^2},$$
thus $\PoA(\GG) \geq \frac{SC(\vec f^*)}{SC(\vec f')} > \frac{1/2M}{2/M^2} = \frac{1}{4M}$, as required. 
\end{proof}

Note that in the above example, the cost for most agents did not increase, whereas the cost of a small fraction of the agents increased by a large factor, which also affected the PoA. We can think about other ways to aggregate the costs of different types, that are less sensitive to such differences. Our next example shows that the equilibrium cost may increase for all types simultaneously. 
\begin{proposition} \label{th:bad_PoA}
For any $M$, there is a PNRG with affine costs $\GG$ such that: $\Psi(\GG)\leq M+2$; and for all agent types, $SC_i(\vec f^*) \geq M \cdot SC^i(\vec f^o)$, where $\vec f^*, \vec f^o$ are the unique equilibrium flow and optimal flow of $\GG$, respectively. In particular, $\PoA(\GG)\geq M$.
\end{proposition} 
\begin{proof}
Consider a graph that is an undirected cycle with $m=M+1$ vertices (such a graph can be implemented by a directed graph as well). For each $i\leq m$, we have $s_i = i$ and $t_i=(i+1)_{\text{mod} m}$. Denote $e_i = (s_i,t_i)$. For type~$i$ agents, the cost function is $c^i_{e^i}(x)=1+x$ and $c^i_e(x)=1$ for all other edges. Suppose $r_i=1$ for all $i$. 

In the optimal flow $\vec f^o$, all type~$i$ agents use edge $e^i$, and thus $SC(\vec f^o,\GG) = \sum_{i\leq m}r_i c^i_{e^i}(1) = (1+\frac1m)\sum_{i\leq m}r_i = m+1$. 
Now consider flow $\vec f^*$, where agent of each type $i$ travel the long path from $s_i$ to $t_i$ using $m-1$ edges. $\vec f^*$ is an equilibrium, since the cost for each agent of type $i$ is $m-1$, whereas the cost on the alternative short path is $c^i_{e^i}(m-1)=m$. Thus the social cost is $SC(\vec f^*,\GG) =   \sum_{i\leq m}r_i (m-1) = m(m-1)$, meaning that $\PoA(\GG) \geq m-1$. 
\end{proof}

Interestingly, the directed version of the graph $G$ in the example has $\PW(G)=m$, and we see no obvious way of using a simpler graph. We conjecture that there is bound on the aggregated social cost similar to that in our main theorem, i.e., that depends on the ``average'' smoothness bounds and on the serial-parallel width of the underlying graph. If true, this would mean that while a large group may have a significant negative externality on a small group of agents from a different type, the opposite is not possible.

\end{document}